\documentclass[a4paper,fleqn]{cas-sc}

\usepackage[authoryear,longnamesfirst]{natbib}
\usepackage{hyperref}
\usepackage{amsmath,color,geometry}
\usepackage{amsmath, amssymb, mathtools}
\usepackage{pstricks}
\usepackage{pst-plot}
\usepackage{parskip,rotating}
\usepackage{pstricks-add}
\usepackage{graphicx}
\usepackage{longtable}
\usepackage{enumitem}
\usepackage{amsmath}
\usepackage{amssymb}
\usepackage{booktabs}
\usepackage{float}
\usepackage{verbatim}
\usepackage{tabularx}
\usepackage{pdflscape}
\usepackage{hyperref}
\usepackage{booktabs}
\usepackage{textcomp}
\usepackage{multirow}
\usepackage{caption}
\usepackage{subfigure}
\usepackage{threeparttable}
\usepackage{etoolbox}
\usepackage{afterpage}
\usepackage{placeins}
\usepackage{setspace}
\AtBeginEnvironment{longtable}{\sffamily}
\usepackage{tikz}
\usepackage{tikz-3dplot}
\usepackage{pgfplots}
\pgfplotsset{compat=1.18}
\usetikzlibrary{patterns, arrows,decorations.pathmorphing,backgrounds,fit,positioning,shapes.symbols,chains,trees,graphs,petri,topaths,shapes.geometric,shapes.arrows,positioning,calc,matrix,decorations.markings,decorations.pathreplacing,intersections,pgfplots.groupplots,automata, fpu, decorations.shapes, intersections,shapes,shapes.multipart, shapes.geometric}
\usetikzlibrary{plotmarks}
\tikzset{
    between/.style args={#1 and #2}{
         at = ($(#1)!0.5!(#2)$)
    }
}
\tikzstyle{decision} = [black, diamond, draw, thick, text width=10em, text badly centered, node distance=3cm, inner sep=0pt, aspect = 2]
\tikzstyle{block} = [rectangle, draw, text width=15em, thick, text centered, minimum height=3em]
\tikzstyle{line} = [draw,thick, -latex']
\tikzstyle{cloud} = [draw, ellipse, node distance=3cm, minimum height=2em]

\usepackage{xcolor}
\usepackage[linesnumbered,ruled,vlined,noend]{algorithm2e}
\usepackage[acronym]{glossaries}
\usepackage[dvipsnames]{xcolor}
\pgfkeys{/tikz/.cd,
	cube top color/.store in=\CubeTopColor,
	cube top color=red!60,
	cube front color/.store in=\CubeFrontColor,
	cube front color=red!30,
	cube side color/.store in=\CubeSideColor,
	cube side color=red!40,
	3d cube color/.code={\colorlet{mycolor}{#1}%
		\tikzset{cube top color=mycolor!60,cube front color=mycolor!30,%
			cube side color=mycolor!40,draw=black, opacity=1}},
		flat cube color/.code={\colorlet{mycolor}{#1}%
		\tikzset{draw=mycolor, cube top color=mycolor!10}}
}

\newacronym{mip}{MIP}{mixed integer programming}
\newacronym{mis}{MIS}{minimal infeasible subset}
\newacronym{ris}{RIS}{reduced infeasible subset}
\newacronym{am}{AM}{additive manufacturing}
\newacronym{bnc}{B\&C}{branch-and-cut}
\newacronym{pbf}{PBF}{powder bed fusion}
\newacronym{2dbpr}{2D-BPR}{two-dimensional bin packing problem with rotation}
\newacronym{2dbpp}{2D-BPP}{two-dimensional bin packing problem}
\newacronym{cp}{CP}{constraint programming}
\newacronym{bpsp}{BPSP}{batch processing scheduling problem}
\newacronym{dffs}{DFFs}{dual feasible functions}
\newacronym{dff}{DFF}{dual feasible function}
\newacronym{1dbpp}{1D-BPP}{one-dimensional bin packing problem}
\newacronym{2dopr}{2D-OPR}{two-dimensional orthogonal packing problem with rotation}
\newacronym{ncbp}{NCBP}{non-contiguous bin packing problem}
\newacronym{csp}{CSP}{constraint satisfaction problem}
\newacronym{1cbp}{1CBP}{bin packing problem with contiguity constraints}
\newacronym{mmim}{MMIM}{minimal meet-in-the-middle patterns}
\newacronym{mhu}{MHU}{Minimum-Height-Up}

\glsdisablehyper

\makeatletter
\pgfdeclareplotmark{half cube*}
{%
	\color{black!50}
	\pgfplots@cube@gethalf@x
	\let\pgfplots@cube@halfx=\pgfmathresult
	\pgfplots@cube@gethalf@y
	\let\pgfplots@cube@halfy=\pgfmathresult
	\pgfplots@cube@gethalf@z
	\let\pgfplots@cube@halfz=\pgfmathresult
	\pgfmathparse{0*\pgfplots@cube@halfz}%
	\let\pgfplots@cube@topz=\pgfmathresult
	\pgfmathparse{-1*\pgfplots@cube@halfz}%
	\let\pgfplots@cube@bottomz=\pgfmathresult
	\pgfplotsifaxissurfaceisforeground{0vv}{%
		\pgfsetfillcolor{\CubeFrontColor}
		\pgfpathmoveto{\pgfplotsqpointxyz{-\pgfplots@cube@halfx}{-\pgfplots@cube@halfy}{\pgfplots@cube@bottomz}}%
		\pgfpathlineto{\pgfplotsqpointxyz{-\pgfplots@cube@halfx}{-\pgfplots@cube@halfy}{\pgfplots@cube@topz}}%
		\pgfpathlineto{\pgfplotsqpointxyz{-\pgfplots@cube@halfx}{ \pgfplots@cube@halfy}{\pgfplots@cube@topz}}%
		\pgfpathlineto{\pgfplotsqpointxyz{-\pgfplots@cube@halfx}{ \pgfplots@cube@halfy}{\pgfplots@cube@bottomz}}%
		\pgfpathclose
		\pgfusepathqfillstroke
	}{%
		\pgfsetfillcolor{\CubeFrontColor}
		\pgfpathmoveto{\pgfplotsqpointxyz{ \pgfplots@cube@halfx}{-\pgfplots@cube@halfy}{\pgfplots@cube@bottomz}}%
		\pgfpathlineto{\pgfplotsqpointxyz{ \pgfplots@cube@halfx}{-\pgfplots@cube@halfy}{\pgfplots@cube@topz}}%
		\pgfpathlineto{\pgfplotsqpointxyz{ \pgfplots@cube@halfx}{ \pgfplots@cube@halfy}{\pgfplots@cube@topz}}%
		\pgfpathlineto{\pgfplotsqpointxyz{ \pgfplots@cube@halfx}{ \pgfplots@cube@halfy}{\pgfplots@cube@bottomz}}%
		\pgfpathclose
		\pgfusepathqfillstroke
	}%
	\pgfplotsifaxissurfaceisforeground{v0v}{%
		\pgfsetfillcolor{\CubeSideColor}
		\pgfpathmoveto{\pgfplotsqpointxyz{-\pgfplots@cube@halfx}{-\pgfplots@cube@halfy}{\pgfplots@cube@bottomz}}%
		\pgfpathlineto{\pgfplotsqpointxyz{-\pgfplots@cube@halfx}{-\pgfplots@cube@halfy}{\pgfplots@cube@topz}}%
		\pgfpathlineto{\pgfplotsqpointxyz{ \pgfplots@cube@halfx}{-\pgfplots@cube@halfy}{\pgfplots@cube@topz}}%
		\pgfpathlineto{\pgfplotsqpointxyz{ \pgfplots@cube@halfx}{-\pgfplots@cube@halfy}{\pgfplots@cube@bottomz}}%
		\pgfpathclose
		\pgfusepathqfillstroke
	}{%
		\pgfsetfillcolor{\CubeSideColor}
		\pgfpathmoveto{\pgfplotsqpointxyz{-\pgfplots@cube@halfx}{ \pgfplots@cube@halfy}{\pgfplots@cube@bottomz}}%
		\pgfpathlineto{\pgfplotsqpointxyz{-\pgfplots@cube@halfx}{ \pgfplots@cube@halfy}{\pgfplots@cube@topz}}%
		\pgfpathlineto{\pgfplotsqpointxyz{ \pgfplots@cube@halfx}{ \pgfplots@cube@halfy}{\pgfplots@cube@topz}}%
		\pgfpathlineto{\pgfplotsqpointxyz{ \pgfplots@cube@halfx}{ \pgfplots@cube@halfy}{\pgfplots@cube@bottomz}}%
		\pgfpathclose
		\pgfusepathqfillstroke
	}%
	\pgfplotsifaxissurfaceisforeground{vv0}{%
		\pgfsetfillcolor{\CubeTopColor}
		\pgfpathmoveto{\pgfplotsqpointxyz{-\pgfplots@cube@halfx}{-\pgfplots@cube@halfy}{\pgfplots@cube@bottomz}}%
		\pgfpathlineto{\pgfplotsqpointxyz{-\pgfplots@cube@halfx}{ \pgfplots@cube@halfy}{\pgfplots@cube@bottomz}}%
		\pgfpathlineto{\pgfplotsqpointxyz{ \pgfplots@cube@halfx}{ \pgfplots@cube@halfy}{\pgfplots@cube@bottomz}}%
		\pgfpathlineto{\pgfplotsqpointxyz{ \pgfplots@cube@halfx}{-\pgfplots@cube@halfy}{\pgfplots@cube@bottomz}}%
		\pgfpathclose
		\pgfusepathqfillstroke
	}{%
		\pgfsetfillcolor{\CubeTopColor}
		\pgfpathmoveto{\pgfplotsqpointxyz{-\pgfplots@cube@halfx}{-\pgfplots@cube@halfy}{\pgfplots@cube@topz}}%
		\pgfpathlineto{\pgfplotsqpointxyz{-\pgfplots@cube@halfx}{ \pgfplots@cube@halfy}{\pgfplots@cube@topz}}%
		\pgfpathlineto{\pgfplotsqpointxyz{ \pgfplots@cube@halfx}{ \pgfplots@cube@halfy}{\pgfplots@cube@topz}}%
		\pgfpathlineto{\pgfplotsqpointxyz{ \pgfplots@cube@halfx}{-\pgfplots@cube@halfy}{\pgfplots@cube@topz}}%
		\pgfpathclose
		\pgfusepathqfillstroke
	}%
}
\makeatother

\newcommand{\Autoref}[1]{%
  \begingroup%
  \def\chapterautorefname{Chapter}%
  \def\sectionautorefname{Section}%
  \def\subsectionautorefname{Subsection}%
  \def\chapterautorefname{Chapter}%
  \def\sectionautorefname{Section}%
  \def\subsectionautorefname{Subsection}%
  \def\subsubsectionautorefname{Subsubsection}%
  \def\paragraphautorefname{Paragraph}%
  \def\tableautorefname{Table}%
  \def\algorithmautorefname{Algorithm}%
  \def\equationautorefname{Equation}%
  \autoref{#1}%
  \endgroup%
}
\SetCommentSty{mycommfont}

\SetKwInput{KwInput}{Input}                
\SetKwInput{KwOutput}{Output}              

\SetInd{0.5em}{0.5em}

\definecolor{gray1}{gray}{0.4}
\definecolor{gray2}{gray}{0.7}
\definecolor{gray3}{gray}{0.9}
\definecolor{color0}{HTML}{B33F62}  
\definecolor{color1}{HTML}{0C855F}  
\definecolor{color2}{HTML}{0C0A3E}  
\definecolor{color3}{HTML}{F3C677}  
\newcolumntype{P}{>{\raggedleft\arraybackslash}X}
\newcolumntype{R}{>{\raggedright\arraybackslash}X}
\newcommand{\ali}[2]{\makebox[#1][l]{#2}}

\def\tsc#1{\csdef{#1}{\textsc{\lowercase{#1}}\xspace}}
\tsc{WGM}
\tsc{QE}

\begin{document}
\newcommand{\BCo}{$\text{B\&C}^{\mathrm{Org}}$}
\newcommand{\BCts}{$\text{B\&C}^{\mathrm{TS}}$}
\newcommand{\Che}{$\text{MIP}^{\mathrm{Che}}$}

\let\WriteBookmarks\relax
\def\floatpagepagefraction{1}
\def\textpagefraction{.001}

\shorttitle{Branch-and-cut for integrated planning in additive manufacturing}    

\shortauthors{Zipfel et al.}  

\title [mode = title]{A new branch-and-cut approach for integrated planning in additive manufacturing}  



%

\author[1]{Benedikt Zipfel}[bioid = 1, orcid=0000-0003-3750-2541]

\cormark[1]


\ead{benedikt.zipfel@tu-dresden.de}


\credit{Conceptualization, Methodology, Software, Validation, Formal analysis, Investigation, Visualization, Writing - Original Draft, Writing - Review \& Editing}

\affiliation[1]{organization={Department of Business Administration, esp. Industrial Management},
            addressline={Helmholtzstraße 10}, 
            city={Dresden},
            postcode={01069}, 
            state={Saxony},
            country={Germany}}

\author[1]{Felix Tamke}[bioid = 1, orcid=0000-0003-1650-8936]


\ead{felix.tamke@tu-dresden.de}


\credit{Conceptualization, Methodology, Validation, Writing - Review \& Editing}

\author[1]{Leopold Kuttner}[bioid = 1, orcid=0000-0002-9359-8835]


\ead{leopold.kuttner@tu-dresden.de}


\credit{Conceptualization, Writing - Review \& Editing}


\cortext[1]{Corresponding author}



\begin{abstract}
In recent years, there has been considerable interest in the transformative potential of \acrfull{am} since it allows for producing highly customizable and complex components while reducing lead times and costs.
The rise of \acrshort{am} for traditional and new business models enforces the need for efficient planning procedures for \acrshort{am} facilities. In this area, the assignment and sequencing of components to be built by an \acrshort{am} machine, also called a 3D printer, is a complex problem joining the nesting and scheduling of parts to be printed. This paper proposes a new branch-and-cut algorithm for integrated planning for unrelated parallel machines. The algorithm is based on combinatorial Benders decomposition: The scheduling problem is considered in the master problem, while the feasibility of a solution is checked in the sub-problem. Current state-of-the-art techniques are extended to solve the orthogonal packing with rotation to speed up the solution of the sub-problem. Extensive computational tests on existing instances and a new benchmark instance set show the algorithm's superior performance compared to an existing integrated mixed-integer programming model.



\end{abstract}


\begin{highlights}
\item We present a new exact algorithm for a scheduling problem in additive manufacturing.
\item We extend state-of-the-art techniques to solve the orthogonal packing with rotation.
\item The branch-and-cut algorithm is vastly superior to an existing integrated model.
\item We propose new benchmark instances for the considered planning problem.
\end{highlights}


\maketitle

\section{Introduction}
\label{sec:Intro}
Initially, \acrfull{am} was adopted to speed up time-to-market or to quickly find solutions from rapid prototyping in the pre-production phase, but it is also increasingly being used for serial production in various industries \citep{Kang2018}. The general technology, also known as 3D printing, can be described as the process of building up parts layer by layer, based on a digital data model \citep{thompson2016design}. \acrshort{am} technologies do not require prior tooling activities \citep{Attaran2017} and are able to process several parts simultaneously in the same production step \citep{Kucukkoc2019}. Because \acrshort{am} enables the building of highly complex parts, it is especially suited to customization purposes \citep{guo2013additive}. While research initially studied the technologies and processes, recent studies increasingly focus on production planning aspects of \acrshort{am} \citep{Oh2020}. The growing interest in production planning is partly driven by new on-demand services, such as Factory-as-a-Service and Production-as-a-Service, that are based on \acrshort{am} technologies (see, e.g., \citet{Kang2018}). However, efficient planning and operations management are also essential for traditional industries shifting from prototypical usage of \acrshort{am} technologies to application in series production.

This study addresses the integrated planning problem on \acrshort{am} machines, which incorporates the nesting of parts into batches and the scheduling of those batches on the available machines \citep{Manco2019}. As the nesting and the scheduling problem directly influence each other \citep{Oh2020}, both sub-problems should be tackled simultaneously in order to obtain adequate and high-quality planning results \citep{Kapadia2019}. The necessity of an integrated solution procedure becomes even more apparent in considerations of unrelated parallel machines with different build spaces and processing times. Therefore, we propose a new integrated exact solution approach to minimize makespan in an \acrshort{am} environment with unrelated parallel machines. The main contributions of this paper are the following:
\begin{itemize}
    \item We present a new \acrfull{bnc} solution procedure to minimize the makespan on unrelated parallel \acrshort{am} machines, which is based on combinatorial Benders decomposition.
    \item We adopt and extend several state-of-the-art techniques from the literature -- such as lower bounding, relaxation, placement point strategies, and constraint programming models -- to solve the sub-problem of the decomposition approach.
    \item We compare our approach with a similar approach from the literature and demonstrate the advantage of our algorithm in terms of computational effort and solution quality by providing more optimally solved instances in fewer computation time.
    \item We propose new benchmark instances for the considered planning problem and demonstrate the superior performance of the proposed approach.
\end{itemize}

\begin{figure}[htb]
    \centering
\begin{tikzpicture}[node distance = 3cm, font = \sffamily]
\node[](1){\resizebox{0.5\textwidth}{!}{\begin{tikzpicture}[>=latex', scale=1, line join=round]
\begin{scope}
\coordinate (Box0-SW) at (0.0,0.0);
\coordinate (Box0-NW) at (0.0,-2.0);
\coordinate (Box0-NE) at (8.0,-2.0);
\coordinate (Box0-SE) at (8.0,0.0);
\draw[solid, line width=0.3mm, color=black] (Box0-SW) -- (Box0-NW) -- (Box0-NE) -- (Box0-SE);
\end{scope}
\begin{scope}
\coordinate (Box1-SW) at (0.0,0.0);
\coordinate (Box1-NW) at (0.0,-1.5);
\coordinate (Box1-NE) at (2.0,-1.5);
\coordinate (Box1-SE) at (2.0,0.0);
\draw[solid, line width=0.3mm, color=black, fill=gray3] (Box1-SW) -- (Box1-NW) -- (Box1-NE) -- (Box1-SE);
\end{scope}
\begin{scope}
\coordinate (Box1-SW) at (1.0,-2.0);
\coordinate (Box1-NW) at (1.0,-1.5);
\draw[solid, line width=0.8mm, color=black] (Box1-SW) -- (Box1-NW);
\end{scope}
\begin{scope}
\coordinate (Box1-SW) at (2.0,-2.0);
\coordinate (Box1-NW) at (2.0,-1.5);
\draw[solid, line width=0.3mm, color=black] (Box1-SW) -- (Box1-NW);
\end{scope}
\begin{scope}
\coordinate (Box1-SW) at (6.0,0.0);
\coordinate (Box1-NW) at (6.0,-1.5);
\coordinate (Box1-NE) at (8.0,-1.5);
\coordinate (Box1-SE) at (8.0,0.0);
\draw[solid, line width=0.3mm, color=black, fill=gray3] (Box1-SW) -- (Box1-NW) -- (Box1-NE) -- (Box1-SE);
\end{scope}
\begin{scope}
\coordinate (Box1-SW) at (7.0,-2.0);
\coordinate (Box1-NW) at (7.0,-1.5);
\draw[solid, line width=0.8mm, color=black] (Box1-SW) -- (Box1-NW);
\end{scope}
\begin{scope}
\coordinate (Box1-SW) at (6.0,-2.0);
\coordinate (Box1-NW) at (6.0,-1.5);
\draw[solid, line width=0.3mm, color=black] (Box1-SW) -- (Box1-NW);
\end{scope}
\begin{scope}
\coordinate (Box1-SW) at (2.0,0.0);
\coordinate (Box1-NW) at (2.0,-1.0);
\coordinate (Box1-NE) at (6.0,-1.0);
\coordinate (Box1-SE) at (6.0,0.0);
\draw[solid, line width=0.3mm, color=black, fill=gray3] (Box1-SW) -- (Box1-NW) -- (Box1-NE) -- (Box1-SE);
\end{scope}
\begin{scope}
\coordinate (Box1-SW) at (4.0,-1.0);
\coordinate (Box1-NW) at (4.0,-2.0);
\draw[solid, line width=0.8mm, color=black] (Box1-SW) -- (Box1-NW);
\end{scope}
\begin{scope}
\coordinate (Box1-SW) at (3.0,0.0);
\coordinate (Box1-NW) at (3.0,-0.975);
\coordinate (Box1-NE) at (5.0,-0.975);
\coordinate (Box1-SE) at (5.0,-0.5);
\coordinate (Box2-SE) at (4.5,-0.5);
\coordinate (Box3-SE) at (4.5,0.0);
\draw[solid, line width=0.05mm, color=gray1, fill=gray1] (Box1-SW) -- (Box1-NW) -- (Box1-NE) -- (Box1-SE) -- (Box2-SE) -- (Box3-SE)-- cycle;
\end{scope}
\node [] (0) at (4.0,-2.0) {};
\node [] (1) at (4.0,-2.5) {};
\draw [->, thick] (0) -- (1);
\node [] (2) at (7.0,-2.0) {};
\node [] (3) at (7.0,-2.5) {};
\draw [->, thick] (3) -- (2);
\node [] (4) at (1.0,-2.0) {};
\node [] (5) at (1.0,-2.5) {};
\draw [->, thick] (5) -- (4);
\node [] (6) at (6.9,0.15) {};
\node [] (7) at (6.4,0.15) {};
\draw [->, thick] (6) -- (7);
\node[color = black] at (1.0, -1.0){\scriptsize Powder};
\node[color = black] at (7.0, -1.0){\scriptsize Powder};
\node[color = gray3] at (4.0, -0.5){\scriptsize Part};
\node[anchor = west, color = black] at (1.0, 0.85){\scriptsize Laser};
\node[anchor = west, align = left, color = black] at (4.35, 1.3){\scriptsize X-Y-Scanning \\[-5pt] \scriptsize Mirrors};
\node[anchor = west, color = black] at (4.0,-1.15){\scriptsize Build Platform};
\node[anchor = west, align = left, color = black] at (6.4,0.6){\scriptsize Powder Leveling \\[-5pt] \scriptsize Roller};
\begin{scope}
\coordinate (Box1-SW) at (1.0,1.0);
\coordinate (Box1-NW) at (1.0,1.5);
\coordinate (Box1-NE) at (3.0,1.5);
\coordinate (Box1-SE) at (3.0,1.0);
\draw[solid, line width=0.3mm, color=black, fill=black] (Box1-SW) -- (Box1-NW) -- (Box1-NE) -- (Box1-SE);
\end{scope}

\node[rectangle, minimum width=0.75cm,  rotate=-45, fill=gray3] at (4.15, 1.25) {};
\node[circle, minimum width=0.1cm, fill=gray2] at (7.0, 0.15) {};
\draw[solid, line width=0.3mm, color=gray2] (3.0, 1.25) -- (4.0, 1.25) -- (4.0, 0.0);

\end{tikzpicture}}};
\end{tikzpicture}
    \caption{Generalized explanation of the print process in \acrshort{pbf} technologies following \citet{gardan2016additive}}
    \label{fig:1}
\end{figure}
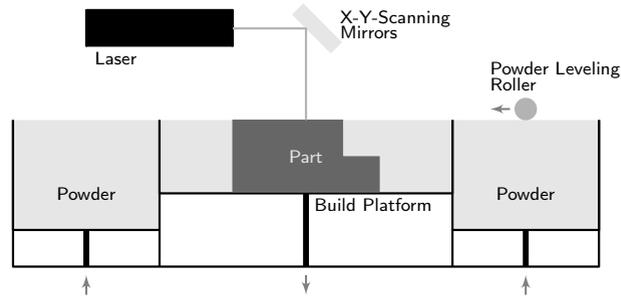

Of the various \acrshort{am} technologies, this study focuses on \acrfull{pbf} techniques, due to their frequent usage in commercial applications \citep{gibson2021additive, Li2017}. \autoref{fig:1} shows a schematic illustration of the general build process of \acrshort{pbf} technologies. A leveling roller spreads powder on the build area; then, a laser beam melts designated areas of the powder to create a layer of the part. When the laser has completed the fusion process on the current layer, the build platform is lowered, and the leveling roller brings out a new layer of powder \citep{gibson2021additive}. Consequently, the processing time depends on the maximum height and the total volume that needs to be printed.

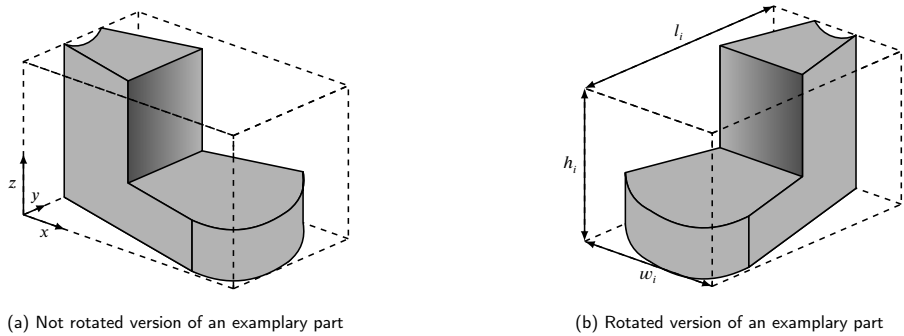
\begin{figure}[htb]
    \centering
    \begin{tikzpicture}
      \node[](1){\resizebox{0.75\textwidth}{!}{\tdplotsetmaincoords{67}{132}

\begin{tikzpicture}
\node [] (2) at (0.0, 0.0) {};
\node [] (3) at (10.0, 0.0) {};
\node [black] (4) at (0.0, -.5) {\sffamily (a) Not rotated version of an examplary part};
\node [black] (5) at (10.0, -.5) {\sffamily (b) Rotated version of an examplary part};
\node [anchor=south] (Boxes-2) at (2) {\begin{tikzpicture}[scale=1,tdplot_main_coords,line join=round] 


\begin{scope}
\coordinate (Box-1-BSW) at (0.0,0.0,0.0);
\coordinate (Box-1-BNW) at (0.0,5.1,0.0);
\coordinate (Box-1-BNE) at (3.1,5.1,0.0);
\coordinate (Box-1-BSE) at (3.1,0.0,0.0);
\coordinate (Hex-B1) at (0.0,4.0,0.0);
\coordinate (Hex-B2) at (1.5,5.0,0.0);
\coordinate (Hex-B3) at (3.0,4.0,0.0);
\coordinate (Hex-B4) at (2.0,0.0,0.0);
\coordinate (Hex-B5) at (1.0,0.0,0.0);

\coordinate (Hex-T1) at (0.0,4.0,1.0);
\coordinate (Hex-T2) at (1.5,5.0,1.0);
\coordinate (Hex-T3) at (3.0,4.0,1.0);
\coordinate (Hex-T4) at (2.0,0.0,3.0);
\coordinate (Hex-T5) at (1.0,0.0,3.0);
\coordinate (Hex-T51) at ($ (Hex-T5) !.5! (0.0,4.0,3.0) $);
\coordinate (Hex-T11) at ($ (Hex-T5) !.5! (0.0,4.0,-1.0) $);
\coordinate (Hex-T41) at ($ (Hex-T4) !.5! (3.0,4.0,3.0) $);
\coordinate (Hex-T31) at ($ (Hex-T4) !.5! (3.0,4.0,-1.0) $);
\coordinate (Box-1-TSW) at (0.0,0.0,3.0);
\coordinate (Box-1-TNW) at (0.0,5.1,3.0);
\coordinate (Box-1-TNE) at (3.1,5.1,3.0);
\coordinate (Box-1-TSE) at (3.1,0.0,3.0);
\draw[dashed, thick, color=black, fill=none] (Box-1-BNE) -- (Box-1-BNW) -- (Box-1-TNW) -- (Box-1-TNE) -- cycle;
\draw[solid, thick, color=black, fill=none] (Hex-B1) to[bend left = 25] (Hex-B2) to[bend left = 25] (Hex-B3) -- (Hex-B4) to[bend right = 45] (Hex-B5) -- cycle;
\draw[dashed, thick, color=black, fill=none] (Box-1-TSW) -- (Box-1-BSW);
\draw[dashed, thick, color=black, fill=none] (Box-1-BSE) -- (Box-1-BSW);
\draw[dashed, thick, color=black, fill=none] (Box-1-BNW) -- (Box-1-BSW);
\draw[solid, thick, color=black, fill=gray!60] (Hex-T1) to[bend left] (Hex-T2) to[bend left] (Hex-T3) -- (Hex-T31) -- (Hex-T41)  -- (Hex-T4) to[bend right = 45] (Hex-T5) -- (Hex-T51) -- (Hex-T11) -- cycle;
\draw[solid, thick, color=black, fill=gray!60] (Hex-T1) to[bend left] (Hex-T2) to[bend left] (Hex-T3) -- (Hex-B3) to[bend right] (Hex-B2) to[bend right] (Hex-B1) -- cycle;
\draw[solid, thick, color=black, fill=gray!60]  (Hex-T3) -- (Hex-T31) -- (Hex-T41)  -- (Hex-T4) -- (Hex-B4) -- (Hex-B3) -- cycle;
\draw[solid, thick, color=black, shading = axis,left color=black!70, right color=gray!60,shading angle=90]  (Hex-T31) -- (Hex-T41)  -- (Hex-T51) -- (Hex-T11) -- cycle;
\draw[dashed, thick, color=black, fill=none] (Box-1-BNE) -- (Box-1-BSE) -- (Box-1-TSE) -- (Box-1-TNE) -- cycle;
\draw[dashed, thick, color=black, fill=none] (Box-1-TNE) -- (Box-1-TNW) -- (Box-1-TSW) -- (Box-1-TSE) -- cycle;

\coordinate (Length) at ($ (Box-1-BSE) !.2! (Box-1-BSW) $);
\coordinate (Width) at ($ (Box-1-BSE) !.2! (Box-1-BNE) $);
\coordinate (Height) at ($ (Box-1-BSE) !.4! (Box-1-TSE) $);
\draw[->, >=latex, solid, thick, color=black, fill=none] (Box-1-BSE) -- node[pos=0.5, above]{$y$} (Length);
\draw[->, >=latex, solid, thick, color=black, fill=none] (Box-1-BSE) -- node[pos=0.5, below]{$x$} (Width);
\draw[->, >=latex, solid, thick, color=black, fill=none] (Box-1-BSE) -- node[pos=0.5, left]{$z$} (Height);

\end{scope}
\end{tikzpicture}
};
\node [anchor=south] (Boxes-3) at (3) {\begin{tikzpicture}[scale=1,tdplot_main_coords,line join=round] 


\begin{scope}
\coordinate (Box-1-BSW) at (0.0,0.0,0.0);
\coordinate (Box-1-BNW) at (0.0,3.1,0.0);
\coordinate (Box-1-BNE) at (5.1,3.1,0.0);
\coordinate (Box-1-BSE) at (5.1,0.0,0.0);
\coordinate (Hex-B1) at (4.0,0.0,0.0);
\coordinate (Hex-B2) at (5.0,1.5,0.0);
\coordinate (Hex-B3) at (4.0,3.0,0.0);
\coordinate (Hex-B4) at (0.0,2.0,0.0);
\coordinate (Hex-B5) at (0.0,1.0,0.0);

\coordinate (Hex-T1) at (4.0,0.0,1.0);
\coordinate (Hex-T2) at (5.0,1.5,1.0);
\coordinate (Hex-T3) at (4.0,3.0,1.0);
\coordinate (Hex-T4) at (0.0,2.0,3.0);
\coordinate (Hex-T5) at (0.0,1.0,3.0);
\coordinate (Hex-T51) at ($ (Hex-T5) !.5! (4.0,0.0,3.0) $);
\coordinate (Hex-T11) at ($ (Hex-T5) !.5! (4.0,0.0,-1.0) $);
\coordinate (Hex-T41) at ($ (Hex-T4) !.5! (4.0,3.0,3.0) $);
\coordinate (Hex-T31) at ($ (Hex-T4) !.5! (4.0,3.0,-1.0) $);
\coordinate (Box-1-TSW) at (0.0,0.0,3.0);
\coordinate (Box-1-TNW) at (0.0,3.1,3.0);
\coordinate (Box-1-TNE) at (5.1,3.1,3.0);
\coordinate (Box-1-TSE) at (5.1,0.0,3.0);
\draw[dashed, thick, color=black, fill=none] (Box-1-BNE) -- (Box-1-BNW) -- (Box-1-TNW) -- (Box-1-TNE) -- cycle;
\draw[solid, thick, color=black, fill=none] (Hex-B1) to[bend right = 25] (Hex-B2) to[bend right = 25] (Hex-B3) -- (Hex-B4) to[bend right = 45] (Hex-B5) -- cycle;
\draw[dashed, thick, color=black, fill=none] (Box-1-TSW) -- (Box-1-BSW);
\draw[dashed, thick, color=black, fill=none] (Box-1-BSE) -- (Box-1-BSW);
\draw[dashed, thick, color=black, fill=none] (Box-1-BNW) -- (Box-1-BSW);
\draw[solid, thick, color=black, fill=gray!60] (Hex-T1) to[bend right] (Hex-T2) to[bend right] (Hex-T3) -- (Hex-T31) -- (Hex-T41)  -- (Hex-T4) to[bend left = 45] (Hex-T5) -- (Hex-T51) -- (Hex-T11) -- cycle;
\draw[solid, thick, color=black, fill=gray!60] (Hex-T1) to[bend right] (Hex-T2) to[bend right] (Hex-T3) -- (Hex-B3) to[bend left] (Hex-B2) to[bend left] (Hex-B1) -- cycle;
\draw[solid, thick, color=black, fill=gray!60]  (Hex-T3) -- (Hex-T31) -- (Hex-T41)  -- (Hex-T4) -- (Hex-B4) -- (Hex-B3) -- cycle;
\draw[solid, thick, color=black, shading = axis,left color=black!70, right color=gray!60,shading angle=270]  (Hex-T31) -- (Hex-T41)  -- (Hex-T51) -- (Hex-T11) -- cycle;
\draw[dashed, thick, color=black, fill=none] (Box-1-BNE) -- (Box-1-BSE) -- (Box-1-TSE) -- (Box-1-TNE) -- cycle;
\draw[dashed, thick, color=black, fill=none] (Box-1-TNE) -- (Box-1-TNW) -- (Box-1-TSW) -- (Box-1-TSE) -- cycle;

\coordinate (Length) at ($ (Box-1-BSE) !.2! (Box-1-BSW) $);
\coordinate (Width) at ($ (Box-1-BSE) !.2! (Box-1-BNE) $);
\coordinate (Height) at ($ (Box-1-BSE) !.4! (Box-1-TSE) $);
\draw[<->, >=latex, solid, thick, color=black, fill=none] (Box-1-TSE) -- node[pos=0.5, above]{$l_i$} (Box-1-TSW);
\draw[<->, >=latex, solid, thick, color=black, fill=none] (Box-1-BSE) -- node[pos=0.5, below]{$w_i$} (Box-1-BNE);
\draw[<->, >=latex, solid, thick, color=black, fill=none] (Box-1-BSE) -- node[pos=0.5, left]{$h_i$} (Box-1-TSE);

\end{scope}
\end{tikzpicture}
};
\end{tikzpicture}}};
    \end{tikzpicture}
    \caption{Illustration of the considered bounding boxes of a part}
    \label{fig:BoundingBox}
\end{figure}

The nesting problem usually takes into account the placement and orientation of parts within the build area. If not all parts fit into a single build area, it is also necessary to group parts into several batches \citep{Oh2020}. In this study, we consider the bounding boxes of each part. This is illustrated in \Autoref{fig:BoundingBox}, in which the cuboid bounding box of a model part is represented by dashed lines.

The terms 'nesting' and 'packing' are sometimes used as synonyms; however, as we understand it, nesting is associated with irregular shapes \citep[see, e.g.,][]{BALDACCI201417}, whereas packing is mostly associated with rectangular shapes. Therefore, as this study considers cuboid bounding boxes with rectangular base areas, we use the term 'packing' in the remainder of this work. Since we do not allow the stacking of parts, each part must be connected to the build plate of a printer. Furthermore, we assume a fixed build orientation of each part, which is predefined in the design process. Parts can only be rotated 90 degrees around the z-axis. This is also illustrated in \Autoref{fig:BoundingBox}, with both rotation variants of a model part. Consequently, the studied problem incorporates both the packing of parts as a \acrfull{2dbpr} and the scheduling of the resulting batches on the available unrelated machines. The machines differ in terms of their processing speeds and build spaces. 

The remainder of this paper is organized as follows. \Autoref{sec:RW} reviews the relevant literature in scheduling problems, packing problems, and planning for additive manufacturing. In \Autoref{sec:PD}, we define the studied problem and formulate a mixed-integer linear programming model; \Autoref{sec:ABCA} presents the structure of the proposed solution approach in detail. We give an overview of the general solution strategy before elaborating on the specific elements of the approach. In our computational studies in \Autoref{sec:CE}, we first analyze data from related literature, after which new benchmark test data is explained and evaluated. Finally, we conclude in \Autoref{sec:C} and give prospects for future research.

\section{Related work}
\label{sec:RW}
We review relevant articles from literature streams that are related to the problem configuration considered here. We first focus on batch scheduling problems; then, the review continues with relevant studies found in the packing and cutting literature. Finally, we present recent papers that explicitly consider packing and scheduling for \acrshort{am}.

\subsection{Scheduling on batch processing machines}
Scheduling on batch processing machines has been studied extensively in the recent decades. Initially, \citet{UZSOY1994} introduced the batch processing scheduling problem (BPSP) for a single batch processing machine; since then, several authors have extended the problem. The most relevant extensions related to the problem examined in this study are non-identical job sizes and unrelated parallel machines. \citet{li2013scheduling} study batch scheduling with these two extensions to minimize makespan, combining different heuristics in two-step approaches and evaluating their performance. More recently, \citet{Arroyo2017b} propose a mathematical programming model for a similar problem, extended by unequal ready times; alongside the exact solution approach, they propose several heuristics to solve the problem. \citet{Arroyo2017a} also address machine-dependent processing times of the parts and different machine capacities. Additionally, a combination of batch scheduling and two-dimensional bin packing is found in \citet{Polyakovskiy2021}, who aim to minimize total weighted earliness and tardiness by introducing different combinations of \acrfull{mip} models and \acrfull{cp} models. 

Similar to the present study, \citet{li2013scheduling} and \citet{Arroyo2017b} take into account parts with different sizes. However, the unrelated machines in those studies differ only based on the processing times, while in our case the machines also have different capacities. In \citet{Arroyo2017a}, the authors extend their previous work and integrate varying machine sizes. All three of these articles consider a one-dimensional packing case, while the present problem assumes placement on a two-dimensional platform. A main difference between the planning problems for batch processing and \acrshort{am} lies in the determination of processing times: in traditional batch scheduling, the processing time of a batch is typically defined by the maximum or the sum of all job processing times, while the processing time in \acrshort{am} depends on the composition of a batch \citep{Alicastro2021}. In contrast to the work of \citet{Polyakovskiy2021}, wherein the number of parts in the batch is relevant for the processing time, in \acrshort{am} the processing times of batches depend on the actual dimensions of the batched parts. Furthermore, our work differs from their work by considering non-guillotine packing.

\subsection{Two-dimensional packing}
\label{subsec:TDP}
For a comprehensive overview of two-dimensional packing literature, we refer to \citet{Iori2021}, who provide an extensive summary of the solution approaches that investigate this problem class. Our study falls under the category of \acrfull{2dbpp} with variable bin sizes and orthogonal rotations. \citet{PISINGER2005154} address \acrshort{2dbpp} with variable bin sizes and propose an exact solution approach based on branch-and-price. Moreover, they present several lower bounds for the problem, including a bound based on Dantzig-Wolfe decomposition. Several publications in the packing literature discuss orthogonal rotation of parts. Here, we give special attention to those articles discussing lower bounding techniques for this problem variant. \citet{DellAmico2002b} present a lower bound for 2D-BPP with rotation (\acrshort{2dbpr}), for which the parts are cut into squares; the new lower bounding technique is embedded in a branch-and-bound algorithm, and its efficiency against the continuous lower bound is demonstrated in a computational study. \citet{Boschetti2003} propose another lower bound by modifying part sizes. A summary of lower bounds for the \acrshort{2dbpr} is presented by \citet{clautiaux2007new}, who also present a new lower bound for the \acrshort{2dbpr} if the considered bin is a square; the authors show the dominance of the newly proposed bound in computational tests. \citet{polyakovskiy2018hybrid} present yet another bound for the \acrshort{2dbpr} based on \acrfull{dffs}, and they show the superior performance of their bound compared to the lower bound of \citet{clautiaux2007new} and the bound resulting from CPLEX. The study of \citet{cote2021combinatorial} solves the \acrshort{2dbpp} with a new branch-and-cut approach based on Benders decomposition. The authors incorporate several state-of-the-art mechanisms and methods from the packing literature to improve the best-known results for the benchmark sets. While \citet{cote2021combinatorial} consider the \acrshort{2dbpp}, in our work, the \acrshort{2dbpr} is a sub-problem of the integrated problem configuration; furthermore, these authors do not consider the rotation of parts. Nevertheless, the ideas presented in that study significantly influenced the investigations in our paper and will therefore be referred to in different sections of this study.

\subsection{Integrated planning for additive manufacturing}
\label{subsec:PfAM}
\subsubsection{Integrated planning with one-dimensional packing}
The number of studies explicitly incorporating packing and scheduling for \acrshort{am} is growing very quickly. \citet{Li2017} introduce a mixed-integer programming model to minimize production costs for the problem of scheduling \acrshort{am} machines; they add tolerances to the production areas of the parts and check whether the summed area of assigned parts is smaller than the available build area on the machine. This \acrfull{1dbpp} constraint is adopted by \citet{Kucukkoc2019}, who studies scheduling in \acrshort{am} for single, identical, and unrelated parallel machines: for each machine environment, the author presents an \acrshort{mip} model that aims to minimize makespan. \cite{TavakkoliMoghaddam2020} tackle makespan and total tardiness simultaneously in a bi-objective approach; in addition, they extend the work of \citet{Kucukkoc2019} by considering material types, which leads to incompatibilities between products. The approach proposed by \citet{Chergui2018} considers customer orders, which are composed of several parts. In their approach, the problem is split into two sub-problems dealing with the assignment of batches and the sequencing on parallel identical machines. \citet{zipfel2021customer} link the extensions of \citet{TavakkoliMoghaddam2020} with the order-based configuration of \citet{Chergui2018} and propose a new MIP formulation. In addition to different material types, the authors also incorporate quality levels for the ordered parts. Elsewhere, a matheuristic for minimizing total weighted tardiness is proposed by \citet{rohaninejad2021scheduling}, who present a combination of a genetic algorithm and a local search based on an \acrshort{mip} model: the machine assignment is addressed in the genetic algorithm, while the MIP-based local search takes care of assigning parts to batches. In \citet{Alicastro2021}, an iterated local search is combined with reinforcement learning; the authors further adopt heuristic approaches from strip packing problems to place the parts within the available batches.

\subsubsection{Integrated planning with two-dimensional packing}
What all mentioned approaches for integrated planning in \acrshort{am} have in common is that they use a \acrshort{1dbpp} constraint instead of explicitly solving the packing problem. This can be a valid assumption if parts are large compared to the build area. However, to meet real-world requirements, the specific dimensions of the parts must be taken into account. \citet{che2021machine} introduce an \acrshort{mip} model that incorporates the scheduling problem, and a two-dimensional packing problem that considers the placement and the orientation of cuboid parts with a rectangular base area on the build area. The orientation variants of each part are predefined by selected rotation movements around the x- and y-axes, while rotations around the z-axis are considered in the model formulation. The model handles orientations around the x- and y-axes using a constraint that allows only one orientation variant of each part to be placed. In addition to the mathematical model, \citet{che2021machine} present a simulated annealing procedure, which they combine with different packing strategies and post-optimization methods; in a comprehensive computational test, they show the efficiency of their developed algorithms. In \citet{HU2022105847}, the aforementioned problem is extended by unequal release dates of parts; the authors adapt the MIP model presented in \citet{che2021machine} and incorporate constraints to account for the extended problem configuration. Furthermore, they use an adaptive large neighborhood search to solve large-scale instances of the problem. \citet{Zhang2020} consider an integrated approach with nesting of irregularly-shaped parts. First, they present a mathematical formulation, before proposing a population-based metaheuristic that integrates the principles of no-fit polygons and inner-fit polygons.

In this study, we examine minimizing makespan on unrelated batch processing machines in \acrshort{am}, where the batching of parts is represented by a two-dimensional bin packing problem with rotation and variably-sized bins. This problem configuration is an extension of the planning situation seen in \citet{Kucukkoc2019} as well as a special case of the problem in \citet{che2021machine} and \citet{HU2022105847}. The studied scheduling sub-problem is a generalization of single batch processing machine scheduling, which has been proven to be NP-hard for minimizing makespan \citep{UZSOY1994}. Furthermore, the packing sub-problem can be reduced to the \acrshort{1dbpp}; therefore, it is also strongly NP-hard. While most existing work on integrated planning for \acrshort{am} with two-dimensional packing focuses on developing heuristic approaches, in this work, we explicitly focus on creating an exact approach. Conceptually, the procedures build on and extends the methodology in the literature of packing problems presented in \Autoref{subsec:TDP}. The next section gives a description of the problem configuration, followed by a mathematical formulation.

\section{Problem description}
\label{sec:PD}

\subsection{Assumptions}

For the studied problem, we make the following assumptions:

\begin{itemize}[label={\footnotesize$\bullet$}]
    \item All information regarding parts and machines is known and deterministic.
    \item We consider 3D printers with \acrshort{pbf} technology as unrelated parallel machines that may vary in terms of their sizes and production parameters. 
    \item All printers are available from the beginning of the planning period and there are no breakdowns of production or malfunctions.
    \item We consider each part using its bounding box and the respective dimensions (see \Autoref{fig:BoundingBox}).
    \item Each part has a predefined build orientation and is only allowed to rotate 90 degrees around the z-axis.
    \item The volume of a part can be smaller than the cuboid volume of the bounding box of its width, length, and height.
    \item The machines are able to build several parts simultaneously in batches, and we do not allow the stacking of parts within the build spaces.
    \item A part needs to be assigned to exactly one batch on one specific machine. 
    \item As soon as the production of a batch starts, no part can be added to or removed from the batch before the printing process has ended.
    \item The maximum number of batches is equal to the total number of parts on each machine, to guarantee feasibility.
\end{itemize}

By considering the bounding boxes of the parts, we aim to ensure consistent part quality by reducing interference of parts with each other. Furthermore, the build orientation of a part has a significant impact on the later quality of the produced component. Therefore, orientation in the build space is often determined in an earlier process step \citep{Kucukkoc2019}. Different materials can be used in PBF technology, and depending on the material and the associated need for additional support structures, the stacking of parts is possible \citep{Bain2019}. In order to attain general applicability, we assume that support structures are required, and thus stacking parts is not permitted. This assumption is important, since it distinguishes between a three-dimensional and a two-dimensional packing. However, we still need to ensure that a part also fits into the build envelope in terms of its height. 
\subsection{Problem definition}
\label{subsec:PD}
\begin{figure}[htb]
    \centering
    \begin{tikzpicture}
    \node[draw, minimum width = \textwidth, minimum height = 7.5cm](0) { };
      \node[](1){\resizebox{\textwidth}{!}{\input{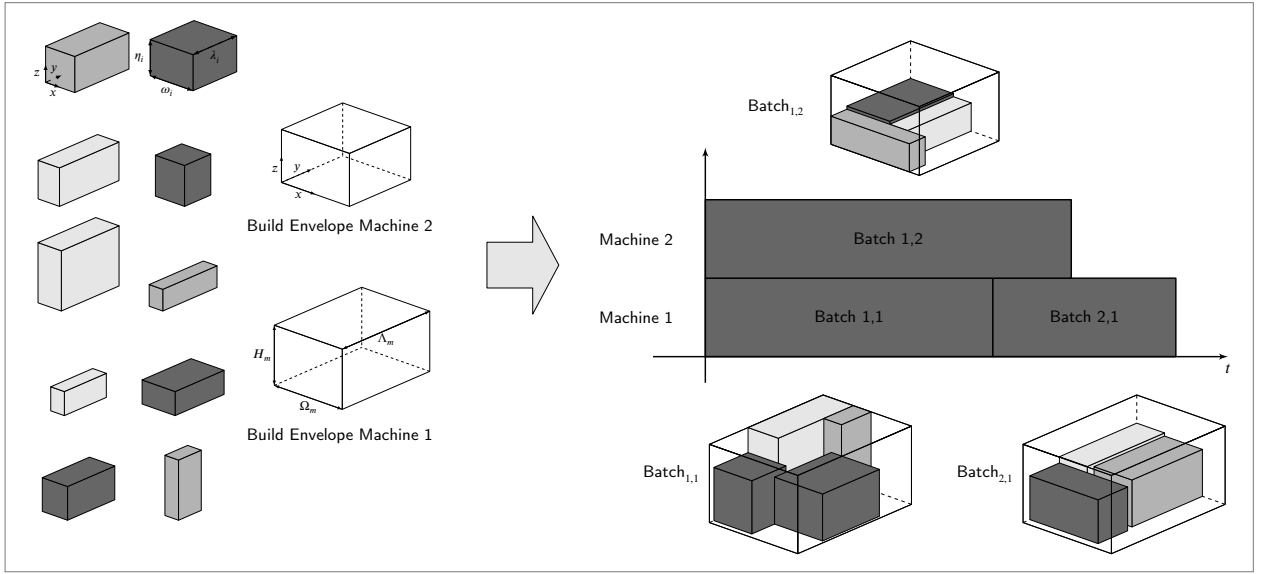}}};
    \end{tikzpicture}
    \caption{Example problem instance and its solution for the considered problem}
    \label{fig:3}
\end{figure}

Let $\mathcal{I} = \{1, 2, ..., I\}$ be a set of parts, each of which is defined by its width $\omega_{i}$, length $\lambda_{i}$, and height $\eta_{i}$, $i \in \mathcal{I}$. Additionally, we are given a set $\mathcal{M} = \{1, 2, ..., M\}$ of machines, which use \acrshort{pbf} technology. The build envelope of an \acrshort{am} machine $m \in \mathcal{M}$ is defined by its width $\Omega_{m}$, length $\Lambda_{m}$, and height $H_{m}$. Besides their variegated build spaces, 3D printers also differ in terms of production times, which can be divided into setup time $T_{m}^{\mathrm{S}}$, recoating time $T_{m}^{\mathrm{R}}$, and scanning time $T_{m}^{\mathrm{L}}$ \citep{Kucukkoc2019}. The recoating time refers to the time to apply a height unit of powder onto the build area, while the scanning time refers to the time the laser needs to build one volume unit. The left side of \Autoref{fig:3} illustrates a model instance of the studied problem with ten parts and two machines, and the right side of the figure presents a possible solution for this. Three batches have been created from the ten parts, based on the available build space of the machines, and those batches are sequenced on the two machines. It should be noted that we numbered the batches in ascending order for each machine -- for example, $\mathrm{Batch}_{1,1}$ is the first batch on the first machine.

In line with the example in \Autoref{fig:3}, the studied problem consists of assigning each part $i \in \mathcal{I}$ to exactly one machine $m \in \mathcal{M}$, packing those parts into a set of batches $\mathcal{B} = \{1, 2, ..., B\}$, placing them within the respective build spaces, and sequencing the resulting batches on the machines. The objective is to minimize the makespan $C_{\mathrm{max}}$ in a machine environment of unrelated parallel machines.

\subsection{Problem formulation}
\label{subsec:PF}
To describe the studied problem mathematically, we modify the formulation of \citet{Kucukkoc2019} by incorporating the ideas of \citet{cote2021combinatorial} for the \acrshort{2dbpp}. Before modeling the problem, we give a complete description of the used notation:

\begin{longtable}{@{}p{0.1\linewidth}p{0.3\linewidth}p{0.08\linewidth}p{0.1\linewidth}p{0.3\linewidth}}
    \toprule
        \multicolumn{2}{l}{\textbf{Classes \& Sets}}&&\\
        \midrule
        $\mathcal{B}$ & \small Set of batches && $\mathcal{I}$ & \small Set of parts\\
        $\mathcal{M}$ & \small Set of machines && $\mathcal{S}$ & \small Generic subset of parts, $S \subseteq \mathcal{I}$\\
         $\mathcal{T}_{m}$ & \small Class of infeasible subsets for part assignments to machine $m$ && $\mathcal{U}$ & Set of tuples $(i, m), i \in \mathcal{I}, m \in \mathcal{M}$,  if $i$ cannot be placed on $m$ \\[4pt]
		\toprule
		\multicolumn{2}{l}{\textbf{Parameters}}&&\\
		\midrule
		$\alpha_{i}$ & \small Production area of part $i \in \mathcal{I}$ && $H_{m}$ & \small Height of machine $m \in \mathcal{M}$\\
    	$\eta_{i}$ & \small Height of part $i \in \mathcal{I}$ && $\Lambda_{m}$ & \small Length of machine $m \in \mathcal{M}$\\
    	$\lambda_{i}$ & \small Length of part $i \in \mathcal{I}$ && $\Omega_{m}$ & \small Width of machine $m \in \mathcal{M}$\\
    	$\upsilon_{i}$ & \small Volume of part $i \in \mathcal{I}$ && $T_m^{\mathrm{L}}$ & \small Scanning time for one volume unit on machine $m \in \mathcal{M}$\\
    	$\omega_{i}$ & \small Width of part $i \in \mathcal{I}$ && $T_m^{\mathrm{R}}$ & \small Recoating time for one height unit on machine $m \in \mathcal{M}$\\
    	$A_{m}$ & \small Area of machine $m \in \mathcal{M}$ &&$T_m^{\mathrm{S}}$ & \small Setup time on machine $m \in \mathcal{M}$\\[4pt]
    	\toprule
		\multicolumn{2}{l}{\textbf{Decision Variables}}&&\\
		\midrule
		$c_{bm}$ & \small Completion time of batch $b \in \mathcal{B}$ on machine $m \in \mathcal{M}$ && $x_{ibm}$ & \small 1, if part $i \in \mathcal{I}$ is assigned to batch $b \in \mathcal{B}$ on machine $m \in \mathcal{M}$\\
    	$h_{bm}$ & \small Height of batch $b \in \mathcal{B}$ on machine $m \in \mathcal{M}$ && $C_{\mathrm{max}}$ & \small Makespan\\
    	$y_{bm}$ & \small 1, if batch $b \in \mathcal{B}$ is processed on machine $m \in \mathcal{M}$ &&& \\[4pt]
    	\bottomrule
	\end{longtable}
\setcounter{table}{0}
With this notation on hand, the studied problem can be stated as an \acrshort{mip} model as follows:

\allowdisplaybreaks
\begin{align}
& C_{\mathrm{max}} \rightarrow \min && \label{E1}\\[2pt]
&\text{s.t.:} && \nonumber \\
&C_{\mathrm{max}} \ge c_{bm} && \forall b \in \mathcal{B}, m \in \mathcal{M}, \label{E2}\\[2pt]
& \sum_{b \in \mathcal{B}}\sum_{m \in \mathcal{M}} x_{ibm} = 1 && \forall i \in \mathcal{I}, \label{E3}\\
& x_{ibm} \le y_{bm} && \forall i \in \mathcal{I}, b \in \mathcal{B}, m \in \mathcal{M}, \label{E4}\\[2pt]
& h_{bm} \ge \eta_{i} \cdot x_{ibm} && \forall i \in \mathcal{I}, b \in \mathcal{B}, m \in \mathcal{M}, \label{E5}\\[2pt]
& h_{bm} \le \max\limits_{i \in \mathcal{I}}\{\eta_{i}\} \cdot y_{bm} && \forall b \in \mathcal{B}, m \in \mathcal{M}, \label{E6}\\[2pt]
& x_{ibm} = 0 && \forall b \in \mathcal{B}, i \in \mathcal{I}, m \in \mathcal{M}, (i, m) \in \mathcal{U}, \label{E15}\\[2pt] 
& c_{1,m} \ge y_{1,m} \cdot T_{m}^{\mathrm{S}} + T_{m}^{\mathrm{L}} \cdot \sum_{i \in \mathcal{I}} (\upsilon_{i} \cdot x_{i,1,m}) + T_{m}^{\mathrm{R}} \cdot h_{1,m} && \forall m \in \mathcal{M}, \label{E8}\\[2pt]
& c_{bm} \ge c_{b-1,m} + y_{bm} \cdot T_{m}^{\mathrm{S}} + T_{m}^{\mathrm{L}} \cdot \sum_{i \in \mathcal{I}} (\upsilon_{i} \cdot x_{ibm}) + T_{m}^{\mathrm{R}} \cdot h_{bm} && \forall b \in \mathcal{B}, b > 1, m \in \mathcal{M}, \label{E9}\\[2pt]
& \sum_{i \in \mathcal{S}} x_{ibm} \le \lvert \mathcal{S} \lvert - 1 && \forall b \in \mathcal{B}, m \in \mathcal{M}, \mathcal{S} \in \mathcal{T}_{m}, \label{E7}\\[2pt]
&  x_{ibm} \in \{0, 1\}, y_{bm} \in \{0, 1\} && \forall i \in \mathcal{I}, b \in \mathcal{B}, m \in \mathcal{M}, \label{E10}\\[2pt]
& C \in \mathbb{R}^{+}, c_{bm} \in \mathbb{R}^{+} && \forall b \in \mathcal{B}, m \in \mathcal{M}, \label{E11}\\[2pt]
& h_{bm} \in \mathbb{N}^{0} && \forall b \in \mathcal{B}, m \in \mathcal{M}. \label{E12}
\end{align}
The objective function defined in \eqref{E1} minimizes the makespan, while constraints \eqref{E2} set the makespan to be greater than or equal to all completion times of batches on the available machines. Constraints \eqref{E3} ensure that each part is assigned to exactly one batch. Constraints \eqref{E4} link variables $x$ and $y$ and establish that batch $b$ is processed on machine $m$ if any part $i$ is assigned to this batch. Inequalities \eqref{E5} and \eqref{E6} limit the height of a batch on a machine $h_{bm}$ to the maximum height of any part $i$ assigned to it. A tuple $(i, m) \in \mathcal{U}$ presents an infeasible assignment of part $i$ to printer $m$. We obtain set $\mathcal{U}$ in preprocessing by checking for all parts whether a part exceeds the width, length, or height of the build space of $m$. In this check, the possibility of rotating a part 90 degrees around the z-axis is also taken into account. Consequently, \eqref{E15} prohibits the assignment of items that have already been proven infeasible for placement on the specific machine. The completion times $c_{bm}$ of batch $b$ on machine $m$ are defined by \eqref{E8} and \eqref{E9}. According to the specifications laid out in \Autoref{subsec:PD}, the completion time of a batch is determined by the setup time, the summed scan time to produce the total volume of the batch, and the time spent to cover the batch height with powder. With \eqref{E9}, we take into account the completion times of previous batches on the same machine. In \eqref{E7}, we use a modification of no-good cuts \citep{martello1990knapsack, cote2021combinatorial} to exclude infeasible batches of the sub-problem that is a \acrfull{2dopr}. By applying these constraints, we forbid the assignment of more than $\lvert \mathcal{S} \lvert$ parts to a batch on machine $m$ for subsets $\mathcal{S} \in \mathcal{T}_{m}$ \citep{cote2021combinatorial}. Variable domains are given in \eqref{E10}--\eqref{E12}.

We adopt the formulation of the unrelated machine environment used in \citet{Kucukkoc2019} for our model. A new feature in the given model is the replacement of area constraints by Benders cuts \eqref{E7}. Because \citet{Kucukkoc2019} focuses explicitly on scheduling for \acrshort{am}, the packing problem is relaxed by implementing area restrictions, allowing assignments only if the summed area of assigned parts is smaller than or equal to the machine's build area. In contrast to this approach, we use the generic formulation adopted from the recent paper of \citet{cote2021combinatorial} on the \acrshort{2dbpp}. Because the evaluation of a given subset $\mathcal{S}$ being part of $\mathcal{T}_{m}$ is NP-complete \citep{CLAUTIAUX20071196}, it is a very challenging task. In the following section, we illustrate our solution approach to the considered problem. In addition, we describe the detailed steps to determine whether a subset $\mathcal{S} \in \mathcal{T}_{m}$ and therefore must be excluded from the solution space.

\section{A branch-and-cut approach}
\label{sec:ABCA}
\subsection{Overall solution procedure}
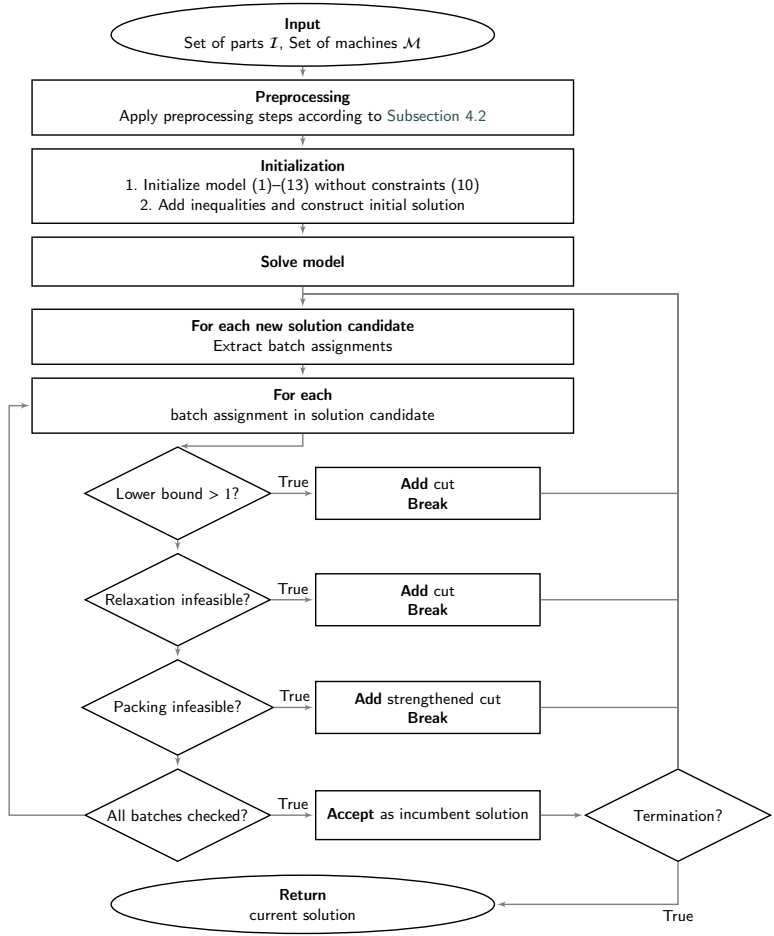
\begin{figure}[htb]
    \centering
    \begin{tikzpicture}
     \node[](1){\resizebox{0.75\textwidth}{!}{\begin{tikzpicture}[auto,
    block_assign/.style ={black, rectangle, draw, thick, fill=white,
      text width=32em, align = center, minimum height=3em, inner sep=6pt},
    block_lost/.style ={black, ellipse, draw, thick, fill=white, 
      text width=16em, text centered},
      line/.style ={draw, thick, -latex', shorten >=1pt}]
    \matrix [ampersand replacement=\&, column sep=18mm,row sep=2.5mm] {
      \& \node [block_lost] (Input) {\textbf{Input}\\ Set of parts $\mathcal{I}$, Set of machines $\mathcal{M}$}; \& \\
      \& \node [block_assign] (Prep){\textbf{Preprocessing}\\Apply preprocessing steps according to \Autoref{subsec:MI}}; \& \\
      \& \node [block_assign] (Init){\textbf{Initialization} \\ 1. Initialize model (1)--(13) without constraints (10) \\ 2. Add inequalities and construct initial solution}; \& \\
      \& \node [block_assign] (Solve) {\textbf{Solve model}}; \&\\
      \& \node [block_assign, yshift = -0.75cm] (Loop1) {\textbf{For each new solution candidate}\\ Extract batch assignments}; \& \\
      \& \node [block_assign] (Loop2) {\textbf{For each} \\ batch assignment in solution candidate}; \& \\
      \& \node[decision, xshift = -2.5cm](Check1){Lower bound $> 1$?}; \node[block_assign, text width = 12.5em, xshift = 2.5cm](Cut1){\textbf{Add} cut\\ \textbf{Break}}; \& \\
      \& \node[decision, xshift = -2.5cm](Check2){Relaxation infeasible?}; \node[block_assign, text width = 12.5em, xshift = 2.5cm](Cut2){\textbf{Add} cut\\ \textbf{Break}}; \& \\
      \& \node[decision, xshift = -2.5cm](Check3){Packing infeasible?}; \node[block_assign, text width = 12.5em, xshift = 2.5cm](Cut3){\textbf{Add} strengthened cut\\ \textbf{Break}}; \& \\
      \& \node[decision, xshift = -2.5cm](Check4){All batches checked?}; \node[block_assign, text width = 12.5em, xshift = 2.5cm](Cut4){\textbf{Accept} as incumbent solution}; \node[decision, xshift = 7.5cm](Check5){Termination?}; \& \\
      \& \node[block_lost, text width = 16em](Finished){\textbf{Return} \\current solution}; \& \\
    };
    \begin{scope}[every path/.style=line]
      \path (Input)   -- (Prep);
      \path (Prep) -- (Init);
      \path (Init) -- (Solve);
      \path (Solve)  -- (Loop1);
      \path (Loop1) -- (Loop2);
      \path (Check1.south) -- (Check2.north);
      \path (Check1.east) -- node[black]{True}(Cut1.west);
      \path (Check2.south) -- (Check3.north);
      \path (Check2.east) -- node[black]{True}(Cut2.west);
      \path (Check3.south) -- (Check4.north);
      \path (Check3.east) -- node[black]{True}(Cut3.west);
      \path (Loop2.south) -- +(0,-0.1) |-  (Check1.north);
      \path [-, shorten >= -0.4pt] (Cut1.east) -| (Check5.north) |- ($(Solve.south)!0.33!(Loop1.north)$);
      \path [-, shorten >= -0.4pt] (Cut2.east) -| (Check5.north) |- ($(Solve.south)!0.33!(Loop1.north)$);
      \path [-, shorten >= -0.4pt] (Cut3.east) -| (Check5.north) |- ($(Solve.south)!0.33!(Loop1.north)$);
      \path (Cut4.east) -- (Check5.west);
      \path [-, shorten >= -0.4pt] (Check5.north) |- ($(Solve.south)!0.33!(Loop1.north)$);
      \path (Check4.east) -- node[black] {True}(Cut4.west);
      \path (Check4.west) -- +(-1.5,0) |- (Loop2);
      \path (Check5.south)  |-  node[black] {True}(Finished.east);
     \end{scope}
 \end{tikzpicture}}};
    \end{tikzpicture}
\caption{General Solution Strategy}
\label{fig:OSS}
\end{figure}

\Autoref{fig:OSS} illustrates the proposed solution method. We consider \eqref{E1}--\eqref{E12} as the master problem, wherein no constraints of type \eqref{E7} are initially imposed. To strengthen the formulated master problem, we conduct several preprocessing steps, apply additional constraints, and construct an initial solution for the problem instance. A detailed description of these improvements can be found in \Autoref{subsec:MI}.

Subsequently, the algorithm solves the problem in a branch-and-cut manner by iteratively adding cuts to the model if necessary. For each solution candidate, we validate the current solution with regard to the feasibility of packed parts in a batch. To this end, we use a step-wise verification mechanism, which consists of lower bounding techniques, relaxation approaches, and the check for packing feasibility by solving the \acrshort{2dopr}. \Autoref{subsec:EVSC} provides detailed explanations of these strategies and their implementation. If an infeasible packing is found, we add a cut to the model and proceed with solving. If no violations can be identified for all used batches, the solution is feasible. The algorithm terminates if the optimal solution is found or if the time limit is reached.

\subsection{Model improvements}
\label{subsec:MI}
\subsubsection{Additional constraints}
We extend the presented model with valid inequalities and fix variables to accelerate the solution process. We allow a total of $B$ batches to be assigned to each available machine. Let $\hat{B}$ denote the number of used batches on a specific machine. Consequently, the total number of possible choices for used batches is $\binom{B}{\hat{B}}$ \citep{che2021machine}. To break this symmetry, we allow a batch $b$ to be opened only if its predecessor $b-1$ on the same machine is also used; we do so by adding inequalities
\begin{align}
& y_{b-1,m} \ge y_{bm} && \forall b \in \mathcal{B},b > 0, m \in \mathcal{M}, \label{additionalE2}
\end{align}
to the model of the master problem. We can further reduce symmetry by prohibiting the assignment of part $i$ to batch $b$ if $b > i$ -- for example, part 1 is only allowed to be assigned to batch 1, while part 2 can be placed in batches 1 and 2, and so on. We integrate this specification for our problem of unrelated machines based on \citet{cote2021combinatorial} with
\begin{align}
& x_{ibm} = 0 && \forall  b \in \mathcal{B}, i \in \mathcal{I}, b > i, m \in \mathcal{M}.  \label{additionalE3}
\end{align}
Constraints \eqref{additionalE4} limit the total area of all parts assigned to a batch $b$ to be less than or equal to the available area $A_{m}$ of machine $m$. These inequalities are typically used to model one-dimensional packing based on area \citep[see, e.g.,][]{Kucukkoc2019,rohaninejad2021scheduling} and to significantly reduce the number of checks in the sub-problem, since the verification mechanisms only need to be called for batches that satisfy the continuous lower bound. 
\begin{align}
& \sum_{i \in \mathcal{I}} x_{ibm} \cdot \alpha_{i} \le A_{m} && \forall b \in \mathcal{B}, m \in \mathcal{M}. \label{additionalE4}
\end{align}

Additionally, we define a set of incompatibilities. Let $\mathcal{Q}_{m}$ be a set of part tuples. A tuple $(i, j) \in \mathcal{Q}_{m}$ establishes that part $i$ and part $j$ cannot be placed in the same batch on machine $m$, since they do not fit together in the build area of $m$. We determine all pairs in $\mathcal{Q}_{m}$ in the preprocessing step and add \eqref{E14} to the model so as to allow only one of the parts in the same batch $b$ on machine $m$. Note that incompatibilities must be defined for each machine individually, due to the machines being unrelated.
\begin{align}
& x_{ibm} + x_{jbm} \le 1 && \forall  b \in \mathcal{B}, i,j \in \mathcal{I}, m \in \mathcal{M}, i < j, (i,j) \in \mathcal{Q}_{m}. \label{E14}
\end{align}

\subsubsection{Reducing the number of available batches}
In \Autoref{subsec:PD}, we assume $I = B$ on each machine in order to ensure feasibility. This assumption leads to many unnecessary decision variables that need to be handled and thus to increased computational effort. We reduce the total batch number $B$ with a heuristic procedure based on the first-fit decreasing height heuristic \citep{chung1982packing, berkey1987two} without losing optimal solutions. Given a set of parts $\mathcal{I}$ and a set of machines $\mathcal{M}$, the algorithm splits the studied problem into individual single machine problems, and for each machine $m \in \mathcal{M}$, we run the same steps, as follows. First, all parts that cannot be placed within the build area of $m$ are discarded. Then, all remaining parts are rotated so that $\omega_{i} \ge \lambda_{i}$ and are sorted by decreasing part width. After that, we fill batches according to first-fit for the respective machine until all parts are assigned. We store the number of needed batches $B_m$ for machine $m$. Finally, $B$ is set to the maximum of all $B_m$ -- i.e., $B = \max\limits_{m \in \mathcal{M}}\{B_m\}$.

\subsubsection{Construction of initial solutions}
To obtain a start solution, we first assign each part to a specific machine. Then, we construct batches in light of the given machine allocation of each part, using the principles of the first fit decreasing heuristic by \citet{johnson1973near} with the extension by shelves \citep{baker1983shelf}. \Autoref{alg:0} summarizes the procedure of finding a start solution to the studied problem. \texttt{StartSolution}$(\mathcal{B},\mathcal{I},\mathcal{M},\mathcal{F})$ creates $\vert \mathcal{F} \vert$ different solutions, where $\mathcal{F}$ represents a set of sorting strategies. In each for-loop, set $\mathcal{I}$ is reordered in accordance with the respective sorting strategy, after which a start solution $s_f$ is generated by applying the described steps of assigning parts to printers and forming batches on these printers. All potential start solutions are stored in set $\mathcal{L}$. \Autoref{alg:0} returns the solution with the lowest makespan, computed with the function $\texttt{GetBestSolution}(\mathcal{L})$. 
\begin{algorithm}[htb]
\DontPrintSemicolon
  \KwInput{Set of batches $\mathcal{B}$; set of parts $\mathcal{I}$, set of machines $\mathcal{M}$, set of sorting strategies $\mathcal{F}$}
  \KwOutput{initial solution $s$}
  \ali{1em}{$\mathcal{L}$} $\xleftarrow{} \, \emptyset$\;
  \ForEach{$f \in \mathcal{F}$}
  {
  sort $\mathcal{I}$ according to $f$\;
  \ali{1em}{$h_f$} $\xleftarrow{} \, \texttt{AssignPartsToMachines}(\mathcal{I};\mathcal{M})$\;
  \ali{1em}{$s_f$} $\xleftarrow{} \, \texttt{ConstructBatches}(\mathcal{B};\mathcal{I};\mathcal{M};h_f)$\;
  \ali{1em}{$\mathcal{L}$} $\xleftarrow{} \, \mathcal{L} \cup \{s_f\}$
  }
  \ali{1em}{$s$} $\xleftarrow{} \quad \texttt{GetBestSolution}(\mathcal{L})$\;
  \Return $s$\;
\caption{\texttt{StartSolution}$(\mathcal{B},\mathcal{I},\mathcal{M},\mathcal{F})$}
\label{alg:0}
\end{algorithm}

The function \texttt{AssignPartsToMachines}$(\mathcal{I}; \mathcal{M})$ aims to balance the workload between the printers: given a set of parts $\mathcal{I}$ and a set of machines $\mathcal{M}$, the algorithm first determines the possible machine assignments for each part using the set of infeasible part-machine assignments $\mathcal{U}$. Part $i$ can only be assigned to machine $m$ if $(i, m) \not \in \mathcal{U}$. After all possible assignments are determined, each part is randomly allocated to one of the possible machines; the resulting assignment permutation $h_f$ represents the starting point for a local search procedure that improves the assignment. A neighbor $h_f^{\prime}$ of a solution $h_f$ is created by changing the machine assignment of one part $i \in \mathcal{I}$; hence, the size of the neighborhood of a solution is $(M - 1) \cdot I$. To evaluate the neighbors, we calculate the maximum workload by determining the processing times on each machine as if all assigned parts fit into one batch and then returning the maximum of all available machines. Thus, we do not consider the dimensions of the parts in this step. The algorithm returns the improved machine assignment permutation $h_f$.

After all parts have been allocated to one of the machines, batches are formed using the function \texttt{ConstructBatches} $(\mathcal{B};\mathcal{I};\mathcal{M}; h_f)$. This procedure begins by assigning the first part to the first batch on the respective machine based on $h_f$. The following parts are successively checked to determine if they fit into an already-existing batch. For checks with only two parts, we conduct a pairwise comparison in terms of rotation around z-axis; if a batch contains more than two parts, feasibility is evaluated using the shelf first-fit decreasing heuristic \citep{baker1983shelf}. If either of the two checks returns a feasible packing of the parts, the considered part is permanently assigned to the batch, and we resume the procedure with the next part. In case of no feasible packing in any existing batches, we open a new batch. This procedure continues until all parts have been placed. The resulting solution $s_f$ is stored in set $\mathcal{L}$ of all constructed solutions. Finally, the best start solution $s$ is selected from among all solutions $\mathcal{L}$.

\subsection{Evaluation of solution candidates}
\label{subsec:EVSC}
Whenever a solution candidate is found for model \eqref{E1}--\eqref{E12}, its feasibility must be verified by checking each of its nonempty batches. Each nonempty batch of a solution contains a specific subset of parts $\mathcal{S} \subseteq \mathcal{I}$. If any check determines that $\mathcal{S}$ is an infeasible batch for its associated machine $m$, we add no-good cuts \eqref{E36} for all batches $b$ of this specific machine $m$.

\begin{align}
 \sum_{i \in \mathcal{S}} x_{ibm} \le \vert \mathcal{S} \vert - 1 && \forall b \in \mathcal{B}. \label{E36}
\end{align}

To guarantee the feasibility or infeasibility of a batch, a \acrshort{2dopr} must be solved. Since this is a computationally challenging problem \citep{cote2021combinatorial}, we want to avoid unnecessary calls to solve the actual \acrshort{2dopr}; instead, we try to prove the infeasibility of a batch as fast as possible. Consequently, we perform the verification of a given batch in multiple steps. First, we use a lower bound technique for the given batch, before we invoke two different relaxation approaches based on \acrshort{dff} and on the \acrfull{ncbp} \citep{boschetti2010exact, cote2014combinatorial}. If we cannot prove infeasibility of the given batch with these steps, we then solve the \acrshort{2dopr} with a \acrshort{cp} model. The different procedures of the verification mechanism are explained in the following subsections.
\subsubsection{Infeasibility by lower bounds}
For each solution candidate, one or more batches are assigned to each machine. Each batch contains a specific subset of parts $\mathcal{S} \subseteq \mathcal{I}$. The first step of our verification mechanism checks whether a batch is feasible by computing a lower bound. We use the bound presented in \cite{DellAmico2002b}, which is based on cutting the parts into squares to make rotations redundant. We denote the considered bound as $LB_{\mathrm{DMV}}$. Given a set $\mathcal{J}$ of square parts after the cutting procedure, subsets of parts $\mathcal{S}_1, \mathcal{S}_2, \mathcal{S}_{23}, \mathcal{S}_3,$ and $\mathcal{S}_4$ are built depending on threshold $q$, s.t. $0 \le q \le \frac{\Lambda}{2}$, and the edge length of each part $j$ given by $\lambda_{j}$, $j \in \mathcal{J}$. These sets are defined by
\begin{alignat}{2}
& \mathcal{S}_{1} && = \Bigg\{ j \in \mathcal{J}: \lambda_{j} > \Omega - q\Bigg\}, \label{E19}\\
& \mathcal{S}_{2} && = \Bigg\{ j \in \mathcal{J}: \Omega - q \ge \lambda_{j} > \frac{\Lambda}{2}\Bigg\}, \label{E20}\\
& \mathcal{S}_{3} && = \Bigg\{ j \in \mathcal{J}: \frac{\Omega}{2} \ge \lambda_{j} > \frac{\Lambda}{2}\Bigg\}, \label{E21}\\
& \mathcal{S}_{23}&& = \Bigg\{ j \in \mathcal{S}_2 \cup \mathcal{S}_3: \lambda_{j} > \Lambda - q\Bigg\}, \label{E22}\\
& \mathcal{S}_{4} && = \Bigg\{ j \in \mathcal{J}: \frac{\Lambda}{2} \ge \lambda_{j} \ge q \Bigg\}. \label{E23}
\end{alignat}
It should be noted that we omit the machine indices of width $\Omega_m$ and length $\Lambda_m$ for simplicity. In the model, the appropriate values for the machines $\Omega_m$ and $\Lambda_m$ are used depending on the machine assignment of a batch. $LB_{\mathrm{DMV}}$ is calculated according to \eqref{E16}--\eqref{E18}, where $\bar{\mathcal{S}}_3 \subseteq \mathcal{S}_{3}$ represents the set of parts in $\mathcal{S}_3$, which can be packed into the batches that pack the parts of $\mathcal{S}_2$ \citep{DellAmico2002b}.
\begin{alignat}{2}
 &\widetilde{LB} &&= \vert \mathcal{S}_{2} \vert + \max\Bigg\{ \Bigg\lceil \frac{\sum_{j \in \mathcal{S}_{3} \backslash \bar{\mathcal{S}}_{3}}\lambda_{j}}{\Omega} \Bigg\rceil , \Bigg\lceil \frac{ \vert \mathcal{S}_{3} \backslash \bar{\mathcal{S}}_{3} \vert }{\big \lfloor \frac{\Omega}{\lfloor \frac{\Lambda}{2} + 1 \rfloor} \big \rfloor} \Bigg\rceil \Bigg\}, \label{E16}\\
& LB(q) &&= \vert \mathcal{S}_{1} \vert + \widetilde{LB} +\max\Bigg\{0, \Bigg\lceil \frac{\sum_{j \in \mathcal{S}_{2}\cup \mathcal{S}_{3}\cup \mathcal{S}_{4}} \lambda_{j}^{2} - (\Omega \cdot \Lambda \cdot \widetilde{LB} - \sum_{j \in \mathcal{S}_{23}} \lambda_{j} \cdot (\Lambda - \lambda_{j}))}{\Omega \cdot \Lambda} \Bigg\rceil \Bigg\}, \label{E17}\\
 & LB_{DMV} &&= \max_{0 \le q \le \frac{\Lambda}{2}} \{ LB(q)\}. \label{E18}
\end{alignat}
Another bound, $LB_{\mathrm{BM}}$, is presented in \citet{Boschetti2003}. In contrast to $LB_{\mathrm{DMV}}$, this bound explicitly takes into account the possible rotation of parts. Due to the time complexity of $O(n^3)$ for $LB_{\mathrm{BM}}$ in comparison to $O(\hat n)$ for $LB_{\mathrm{DMV}}$ with $n = \vert \mathcal{S} \vert, \hat n = \vert \mathcal{J} \vert$, we decided to only use $LB_{\mathrm{DMV}}$ \citep{Boschetti2003,DellAmico2002b}.
If the resulting value for $LB_{\mathrm{DMV}}$ is greater than 1, we have found an infeasible assignment $\mathcal{S}$ and therefore add a no-good cut \eqref{E36} to the master problem \eqref{E1}--\eqref{E12}. Otherwise, the algorithm continues with the next feasibility check.

\subsubsection{Infeasibility by orthogonal relaxation}
In the second step, we try to prove the infeasibility of subset $\mathcal{S}$ using two different relaxation procedures. First, we consider a \acrfull{csp} with feasibility constraints built on \acrshort{dff} \citep[see, e.g.,][]{polyakovskiy2018hybrid}. A \acrshort{dff} represents a function $u:$ $[0, 1] \rightarrow [0, 1]$, in which for any set $\mathcal{K}$ of non-negative real numbers, there is a relation in accordance with inequality \eqref{E52} \citep{fekete2004general}.
\begin{align}
 \sum_{\kappa \in \mathcal{K}} \kappa \le 1 \Longrightarrow \sum_{\kappa \in \mathcal{K}} u(\kappa) \le 1. &&  \label{E52}
\end{align}
Given two \acrshort{dffs} $u_1$ and $u_2$, we are able to transform the scaled width $w'_i$ and length $l'_i$ of part $i \in \mathcal{S}$ into $(u_1(w'_i), u_2(l'_i))$, where $w'_i = w_i/W$ and $l'_i = l_i/L$.
Let $\alpha_{ci}$ be a scaled area, where $\alpha_{ci} = u_1(w'_i) \cdot u_2(l'_i)$. A feasible packing is only possible if the sum of all scaled areas is less than or equal to one \citep{polyakovskiy2018hybrid}, i.e.,
\begin{align}
 \sum_{i \in \mathcal{S}} \alpha_{ci} \le 1. &&  \label{E53}
\end{align}

Let $\mathcal{S}^{\mathrm{E}} = \{1, ..., i, ..., n, n + 1, ..., 2n\}$ be the set of parts in $\mathcal{S}$ augmented by the identical parts rotated by $90$ degrees $(n + 1, ..., 2n)$. The decision variable $d_i$, $i \in \mathcal{S}^{\mathrm{E}}$, indicates whether part $i$ is packed in the batch or not. In addition, $\mathcal{C}$ denotes the set of combined \acrshort{dffs} $u_1$ and $u_2$. Using this notation, the \acrshort{csp} can be defined as:
\begin{align}
  & \sum_{i \in \mathcal{S}} \left ( \alpha_{ci} \cdot d_{i} + \alpha_{c,n + i} \cdot d_{n + i}  \right ) \le 1 && \forall c \in \mathcal{C}, \label{E54} \\[2pt]
  & d_{i} +  d_{n + i}  = 1 && \forall i \in \mathcal{S}, \label{E55} \\[2pt]
  & d_i \in \{0, 1\} && \forall i \in \mathcal{S}^{\mathrm{E}}. \label{E56}
\end{align}

We apply the functions $u^{(\epsilon)}$, $U^{(\epsilon)}$, and $\phi^{(\epsilon)}$ as presented in \citet{fekete2004general}, where $\epsilon = \{p, q\}$, and we define the same combinations of these functions that the authors use for their bound $L_{2d}$. Additionally, we apply the combinations of \acrshort{dffs} from \citet{fekete2007exact} that are used for the check on orthogonal packings. If the model cannot be solved, we have found an infeasible subset $\mathcal{S}$; in this case, we add another cut \eqref{E36} to the model and thus reject the solution candidate.

For the second relaxation procedure, we consider the adjusted version of the \acrfull{1cbp}. In the \acrshort{1cbp}, each item $i$ is cut into $\omega_{i}$ slices of length $\lambda_{i}$ and width 1, and the bin size transforms into length $\Lambda$ and width 1. The goal is to pack all slices into the minimum number of bins, under the condition that all slices of the same part $i$ must be packed side-by-side with each other \citep{cote2014combinatorial}. This condition is omitted for the \acrshort{ncbp}, which is also known as the bar relaxation \citep{belov2013conservative}. To prohibit the stacking of slices of the same part $i \in \mathcal{I}$, we require that at most one slice of each part is packed into the same bin. 

The \acrshort{ncbp} can be described as follows.
A pattern $t$ identifies a subset of parts whose summed length is less than or equal to the machine's length. Originally, the pattern is described by an array $(\sigma_{1t}, ..., \sigma_{it}, ..., \sigma_{nt})^{T}$, where $\sigma_{it}$ is 1 if part $i$ is in the pattern, and 0 otherwise \citep{cote2014combinatorial}. To account for the possible rotation of a part, we additionally allow $\sigma_{it}$ to take the value of ${\omega_i}/{\lambda_i}$ if the part is rotated in the pattern. Let $\mathcal{T}$ be the set of all currently available patterns that include at least one slice of an item, and let integer variable $z_t$ denote how often a pattern $t$ is used in the solution. Then, in accordance with \citet{cote2014combinatorial}, the \acrshort{ncbp} can be modeled as:
\begin{align}
& \sum_{t \in \mathcal{T}} z_t \rightarrow \min && \label{E29}\\[2pt]
&\text{s.t.:} && \nonumber \\
&\sum_{t \in \mathcal{T}}\sigma_{it} \cdot z_{t} \ge \omega_{i} && \forall i \in \mathcal{S}, \label{E30}\\[2pt]
& z_t \in  \mathbb{N}_{\geq 0} && \forall t \in \mathcal{T}. \label{E31}
\end{align}

In \eqref{E29}, we minimize the total number of used patterns. Constraints \eqref{E30} ensure that the total number of appearances of each part $i$ in all used patterns is greater than or equal to the part's width $\omega_{i}$. \eqref{E31} defines the domain of the decision variable $z_t$. Since the \acrshort{ncbp} is an NP-hard problem, we follow \citet{cote2014combinatorial} and only consider the continuous relaxation of \eqref{E29}--\eqref{E31}, for which we use the column generation approach presented by \citet{gilmore1961linear} for the cutting stock problem.
$\mathcal{T}$ is initialized with $n$ patterns, each of which incorporates a slice of one individual part $i$. To take into account the restriction that each pattern $t$ can include only one slice of a part, as well as the possibility to rotate the parts, we adjust the sub-problem of the original column generation approach.

Let $\mathcal{S}^{\mathrm{E}} = \{1, ..., i, ..., n, n + 1, ..., 2n\}$ again be the set of parts in $\mathcal{S}$ augmented by their rotated versions. The binary variable $v_i$ is 1 if a part $i$ is in the solution of the sub-problem, and 0 otherwise. Given the dual values $\delta_i$ of \eqref{E30}, we define the sub-problem as:

\begin{align}
& 1 - \bigg ( \sum_{i = 1}^{n} \delta_i \cdot v_i + \sum_{i = n + 1}^{2n} \frac{\lambda_i}{\omega_i} \cdot \delta_i \cdot v_i \bigg ) \rightarrow \min && \label{E32}\\[2pt]
&\text{s.t.:} && \nonumber \\
&\sum_{i \in \mathcal{S}^{\mathrm{E}}} v_{i} \lambda_{i} \le L, && \label{E33}\\[2pt]
& v_{i} + v_{i + n} \le 1 && \forall i \in \mathcal{S}, \label{E34}\\[2pt]
& v_i \in \{0, 1\} && \forall i \in \mathcal{S}^{\mathrm{E}}. \label{E35}
\end{align}
Similar to other column generation techniques, we seek to find a new pattern to add to the master problem, which we do by finding a combination of part slices that minimizes the objective function \eqref{E32}. To account for the rotation of part $i + n$, the dual value $\delta_i$ of part $i$ is multiplied by the ratio ${\lambda_i}/{\omega_i}$ in the objective function. We ensure that the summed length of slices does not exceed the length of the machine with \eqref{E33}. Constraints \eqref{E34} enforce that only one rotation for each part is used. Whenever the sub-problem is solved, we extract the found solution and create a new pattern to add to the master problem. As stated above, if a part $i$ appears in the solution with its original orientation, we add 1 to the new pattern; in cases in which part $i$ has been rotated, the value of this part in the considered pattern will be ${\omega_i}/{\lambda_i}$.

We continue the search for new patterns until no better solution can be found. In this case, we compare the continuous lower bound of model \eqref{E29}--\eqref{E31} with the width of the machine to which the examined batch has been assigned. If the value of the lower bound is larger than the machine's width, we have found an infeasible subset $\mathcal{S}$; at this point, we add a cut of type \eqref{E36} and continue the solution process of model \eqref{E1}--\eqref{E12}.

\subsubsection{Infeasibility by orthogonal packing}
\label{CIS}
In cases in which the above-mentioned verification steps are not able to prove the infeasibility of a batch, we solve the \acrshort{2dopr} with \acrshort{cp} as an exact method to ensure the infeasibility or feasibility of a batch. \acrshort{cp} has already been proven to be suitable for two-dimensional orthogonal packing in several publications (see, e.g., \citet{martello2007algorithm,pisinger2007using,clautiaux2008new}). Recently, \citet{Polyakovskiy2021} use CP to solve the \acrshort{2dopr} sub-problem in a similar problem configuration.

To model the \acrshort{2dopr} as a \acrshort{csp}, we define variables $x^{\mathrm{S}}_i$, $x^{\mathrm{E}}_i$, $y^{\mathrm{S}}_i$, and $y^{\mathrm{E}}_i$ to describe the start and end positions of a part $i$ on the x-axis and on the y-axis, respectively. The variables $w_i$ and $l_i$ represent the width and length of a part $i$. We use the interval variables
\begin{align}
& interval^{x}_i  \xrightarrow{} x^{\mathrm{S}}_i +  w_i = x^{\mathrm{E}}_i  && \forall i \in \mathcal{S}, \label{E37}\\[2pt]
& interval^{y}_i \xrightarrow{} y^{\mathrm{S}}_i + l_i = y^{\mathrm{E}}_i && \forall i \in \mathcal{S}, \label{E38}
\end{align}
to link the start and end points of part $i$ on an axis with its respective dimension. 
Furthermore, the binary variable $e_{ir}$ is 1 if part $i$ is placed with rotation variant $r \in \mathcal{R}_i$, and 0 otherwise. 
Finally, $\omega_{ir}$ and $\lambda_{ir}$ denote the width and the length, respectively, of part $i$ in rotation $r$. The \acrshort{cp} model for the \acrshort{csp} can be formulated as follows: 
\begin{align}
& \sum_{r \in \mathcal{R}_i} e_{ir} = 1  && \forall i \in \mathcal{S}, \label{E41}\\[2pt]
& \textit{NoOverlap2D}(\{interval^{x}_i: i\in \mathcal{S}\}, \{interval^{y}_i: i\in \mathcal{S}\}), && \label{E42}\\[2pt]
& e_{ir} = 1 \Rightarrow w_i = \omega_{ir} \land l_i = \lambda_{ir} && \forall i \in \mathcal{S}, r \in \mathcal{R}_i. \label{E43}
\end{align}
Constraints \eqref{E41} ensure that exactly one orientation for each part is chosen. Constraint \eqref{E42} is a global constraint and guarantees that all rectangles are non-overlapping. The rectangle associated with part $i$ is defined by the combination of interval variables $interval^{x}_i$ and $interval^{y}_i$. Constraints \eqref{E43} determine the width and length of part $i$ based on the selected rotation variant $r$.
It should be noted that due to the consideration of rectangles and the possibility of rotating a part 90 degrees, the number of possible rotation variants is one if the part is represented as a square, and two otherwise.

In the following, we discuss the domains of the variables. Since there are at most two different rotation variants, the variables $w_i$ and $l_i$ can only take the width and length of the rotated or non-rotated part. Hence, the domains are defined according to
\begin{align}
& w_i \in \bigg \{\min_{r \in \mathcal{R}_i}\{\omega_{iv}\};\max_{r \in \mathcal{R}_i}\{\omega_{iv}\} \bigg \}  && \forall i \in \mathcal{S}, \label{E45}\\[2pt]
& l_i \in \bigg \{\min_{r \in \mathcal{R}_i}\{\lambda_{iv}\};\max_{r \in \mathcal{R}_i}\{\lambda_{iv}\} \bigg \}  && \forall i \in \mathcal{S}. \label{E46}
\end{align}

For the positioning variables $x^{\mathrm{S}}_i$, $x^{\mathrm{E}}_i$, $y^{\mathrm{S}}_i$, and $y^{\mathrm{E}}_i$, we reduce the solution space of the \acrshort{2dopr} by using the principle of \acrfull{mmim} introduced by \citet{cote2018meet}. To respect the rotation of parts, we adjust the procedures \texttt{MinimalMIMSet}$(\mathcal{I};\Omega)$ and \texttt{NormalPatterns}$(\mathcal{I};\Omega)$ presented by these authors. By adding another loop for rotations $r \in \mathcal{R}_i$ in line 4, \Autoref{Al3} respects rotations in normal patterns. The resulting procedure is illustrated as  \texttt{AdjustedNormalPatterns}$(\mathcal{I};\Omega)$ in \Autoref{Al3}. For \texttt{MinimalMIMSet}$(\mathcal{I};\Omega)$, this is done by invoking \texttt{AdjustedNormalPatterns}$(\mathcal{S}\backslash {i}; \Omega - min_{r \in \mathcal{R}_i}\{\omega_{ir}\})$ taking into account the rotations on filling the arrays $T^{\mathrm{left}}, T^{\mathrm{right}}$, and defining the minimal set of placement points for part $i \in \mathcal{I}$. We display the pseudocode for \texttt{AdustedMinimalMIMSet}$(\mathcal{I};\Omega)$ in \Autoref{Al4}. After the sets of placement points $\mathcal{P}_i^{\mathrm{min}, \mathrm{x}}$ and  $\mathcal{P}_i^{\mathrm{min}, \mathrm{y}}$ are determined, the variable domains are defined as: 
\begin{align}
& x^{\mathrm{S}}_i  = &&\bigg\{p : p \in \mathcal{P}_i^{\mathrm{min}, \mathrm{x}} \bigg\} && \forall i \in \mathcal{S}, \label{E47}\\[2pt]
& x^{\mathrm{E}}_i = &&\bigg\{ p + \omega_{ir} : p \in \mathcal{P}_i^{\mathrm{min}, \mathrm{x}}, r \in \mathcal{R}_i \hspace{2pt} \vert \hspace{2pt} p + \omega_{ir} \le \Omega \bigg\} && \forall i \in \mathcal{S}, \label{E48}\\[2pt]
& y^{\mathrm{S}}_i = &&\bigg\{p : p \in \mathcal{P}_i^{\mathrm{min}, \mathrm{y}} \bigg\} && \forall i \in \mathcal{S}, \label{E49}\\[2pt]
& y^{\mathrm{E}}_i = &&\bigg\{p + \lambda_{ir}: p \in \mathcal{P}_i^{\mathrm{min}, \mathrm{y}}, r \in \mathcal{R}_i \hspace{2pt} \vert \hspace{2pt}  p + \lambda_{iv} \le \Lambda \bigg\} && \forall i \in \mathcal{S}. \label{E50}
\end{align}

\begin{algorithm}[htb]
    \caption{\texttt{AdjustedNormalPatterns}$(\mathcal{I};\Omega)$ in accordance with \citet{cote2018meet}}
    \label{Al3}
    \DontPrintSemicolon
    \KwInput{set of parts $\mathcal{I}$, machine width $\Omega$}
    \KwOutput{set of placement points $\mathcal{N}_0$ according to normal pattern principle}
    $T \xleftarrow{} [0 \text{ to } \Omega]$: all entries initialized as $0$\;
    $T[0] \xleftarrow{} 1$\;
    \ForEach{\upshape $i \in \mathcal{I}$}
    {
        \ForEach{\upshape $r \in \mathcal{R}_i$}
        {
            \ForEach{$p = \Omega - \omega_{ir} \text{ to } 0$}
            {
            \If{$T[p] = 1$}{$T[p + \omega_{ir}] \xleftarrow{} 1$}
            }
        }
    }
    $\mathcal{N}_0 \xleftarrow{} \, \emptyset$\;
    \ForEach{$p = \Omega \text{ to } 0$}
    {
    \If{$T[p] = 1$}{$\mathcal{N}_0 \xleftarrow{} \mathcal{N}_0 \cup \{p\}$}
    }
    \Return $\mathcal{N}_0$\;
\end{algorithm}

\begin{algorithm}[htb]
    \caption{\texttt{AdjustedMinimalMIMSet}$(\mathcal{I};\Omega)$  in accordance with \citet{cote2018meet}}
    \label{Al4}
    \DontPrintSemicolon
    \KwInput{set of parts $\mathcal{I}$, machine width $\Omega$}
    \KwOutput{minimal set $\mathcal{P}$ of MIM placement points}
    $T^{\mathrm{left}}, T^{\mathrm{right}} \xleftarrow{} [0 \text{ to } W]$: all entries initialized as $0$\;
    \ForEach{\upshape $i \in \mathcal{I}$}
    {
        $\mathcal{V}_i \xleftarrow{} \, $ \texttt{AdjustedNormalPatterns}$(\mathcal{S}\backslash {i}; \Omega - min_{r \in \mathcal{R}_i}\{\omega_{ir}\})$\;
        \ForEach{$p \in \mathcal{V}_{i}$}
        {
        $T^{\mathrm{left}}[p] = 1$\;
        \ForEach{$r \in \mathcal{R}_{i}$}
        {
        \If{$\omega_{iv} < \Omega$}
        {
        $T^{\mathrm{right}}[\Omega - \omega_{ir} - p]  \xleftarrow{} \, T^{\mathrm{right}}[\Omega - \omega_{ir} - p] + 1$\;
        }
        }
        }
    }
    \ForEach{$p = 1$ to $\Omega$}
    {
    $T^{\mathrm{left}}[p] \xleftarrow{} \,T^{\mathrm{left}}[p]  +T^{\mathrm{left}}[p - 1]$\;
    $T^{\mathrm{right}}[\Omega - p] \xleftarrow{} \,T^{\mathrm{right}}[\Omega - p]  +T^{\mathrm{right}}[\Omega - (p - 1)]$\;
    }
    $t^{\mathrm{min}} \xleftarrow{} 1$\;
    $\min$ $\xleftarrow{} \, T^{\mathrm{left}}[0] + T^{\mathrm{right}}[1]$\;
    \ForEach{$p = 2$ to $\Omega$}
    {
        \If{$T^{\mathrm{left}}[p - 1] + T^{\mathrm{right}}[p] < \min$}
        {
        $\min$ $\xleftarrow{} T^{\mathrm{left}}[p - 1] + T^{\mathrm{right}}[p]$\;
        $t^{\mathrm{min}} \xleftarrow{} p$\;
        }
    }
    $\mathcal{P} \xleftarrow{} \emptyset$\;
    \ForEach{$i \in \mathcal{I}$}
    {
    $\mathcal{P}_i \xleftarrow{}\emptyset$\;
        \ForEach{$p \in \mathcal{V}_i$}
        {
            \If{$p < t^{\mathrm{min}}$}
            {$\mathcal{P}_i \xleftarrow{} \mathcal{P}_i \cup \{p\}$}
            \ForEach{$r \in \mathcal{R}_i$}
            {
                \If{$\Omega - \omega_{ir} - p \geq t^{\mathrm{min}}$}
                {$\mathcal{P}_i \xleftarrow{} \mathcal{P}_i \cup \{\Omega - \omega_{ir} - p\}$}
            }
        }
     {$\mathcal{P} \xleftarrow{} \, \mathcal{P} \cup \mathcal{P}_i$}
    }
    \Return $\mathcal{M}$\;
\end{algorithm}

If the \acrshort{csp} \eqref{E41}--\eqref{E43} ascertains the infeasibility of $\mathcal{S}$, we add another no-good cut. As noted by \citet{cote2021combinatorial}, these cuts are usually weak for large sets $\mathcal{S}$; thus, we prefer finding a \acrfull{mis} $\mathcal{S}^\prime \subseteq \mathcal{S}$ that cannot be packed onto machine $m$. Using $\mathcal{S}^\prime$ instead of $\mathcal{S}$ in no-good cuts \eqref{E36} would lead to a strengthened cut.

However, finding an \acrshort{mis} is a challenging task. Therefore, we apply the heuristic approach from \citet{cote2021combinatorial} of finding a \acrfull{ris}. First, the parts in $\mathcal{S}$ are sorted by ascending area. Then, we iteratively remove the smallest part from the subset and invoke model \eqref{E41}--\eqref{E43} with the reduced set $\mathcal{S}^\prime$ and a time limit. The procedure continues as long as infeasibility can be proved for $\mathcal{S}^\prime$. If infeasibility cannot be proven within the time limit or if the considered subset is feasible, we reinsert the last removed part $i$ into $\mathcal{S}^\prime$ and use $\mathcal{S}^\prime$ to generate a new cut.

\section{Computational experiments}
\label{sec:CE}
In this section, we analyze the performance of the proposed \acrshort{bnc} approach. We consider two variants of the algorithm: the first is the original approach {\BCo}, described in the previous section; the second is a two-step procedure called {\BCts}. In the first step, we try to improve the initial solution by invoking {\BCo} with a more restricted version of constraint \eqref{additionalE4}, such that the summed part areas in a batch are at most 90\% of the respective machine area. In the second step, we use the improved initial solution as a start solution for the original approach with 100\% occupation of machine areas. With this adjusted \acrshort{bnc}, we attempt to avoid getting stuck in checking very dense packings, which are not likely to be part of an optimal solution in the unrestricted version. We compare both variants to the \acrshort{mip} model presented by \citet{che2021machine}, since this is the only other exact solution approach addressing rectangular shapes and unrelated machines; we denote this model as {\Che}. During preliminary testing, we observed some irregularities in the \acrshort{mip} model, as some packing problems with a known feasible solution were declared infeasible. After thorough investigation, we solved these issues by correcting the Big-M values used in the model. The {\Che} model and a comprehensive explanation of the changes can be found in the supplementary material to this study. 

The \acrshort{bnc} approaches have been implemented with \textsc{Gurobi} and \textsc{Google OR-Tools} in \textsc{Python}. We use the former to model the master problem and the orthogonal relaxation procedures, while we use the latter to implement the \acrshort{cp} model of the \acrshort{2dopr}. The {\Che} is also modeled using \textsc{Python} and \textsc{Gurobi}.

We evaluate the performance of the approaches using two different data sets. First, we use the data given in \citet{che2021machine}, which is available online \citep{che2021machine}; then, we analyze the given data set, point out some drawbacks, and propose a new comprehensive test data set. This data set is used to analyze the solution approaches in more detail.

The study is conducted on an AMD EPYC 7513 with 3.0 GHz clock speed and 64 GB RAM. For the computational tests, the maximum number of threads is set to eight, and an overall time limit of 3,600 seconds is defined for the adjusted {{\Che}} model and both \acrshort{bnc} variants. To comply with the overall time limit in the \acrshort{bnc} approaches, we dynamically set the time limit of the \acrshort{cp} model in the feasibility check of a batch to the residual runtime (3,600\,sec minus the elapsed time). \textsc{Gurobi}'s relative MIP optimality gap is set to $7 \cdot 10^{-6}$. The evaluation of a potential cut strengthening in \acrshort{bnc} is restricted to two seconds. With regard to the {\BCts}, the maximum time spent in the restricted version of the master problem is set to 10\% of the overall computation time (i.e., 360 sec). All test runs are repeated three times with different seed values.

\subsection{Comparative tests on existing test data}
\label{CTOETD}
\subsubsection{Instance data}
As stated in \Autoref{sec:RW}, \citet{che2021machine} study a problem configuration for packing and scheduling in \acrshort{am} that is similar to the problem in this paper. In addition to rotation around the z-axis, the authors also allow a subset of orientations around the x- and y-axes. \Autoref{tab:0} presents the given data for an example part. Each part is defined with up to three different orientations, resulting in different parameter values for its width, length, and height, as well as for the needed support structure. The original part data is collected from the website Thingiverse (\url{www.thingiverse.com}), which provides user-produced design data. The authors use four different machine types based on the parameters of \citet{Kucukkoc2019} and \citet{Li2017}; the types differ in the dimensions of the building platform, setup times, and recoating times $T_m^{\mathrm{R}}$ (see \autoref{tab:-2}). \citet{che2021machine} define seven classes of instances, ranging from 20 parts and two machines to 500 parts and seven machines. We focus on the instance classes with 20 and 50 parts (Class 1 and 2), each containing 20 instances, since \citet{che2021machine} have solved them with their integrated \acrshort{mip} model. The remaining instance classes have only been solved with a simulated annealing approach. 

\renewcommand{\arraystretch}{1.05}
\begin{table}[!htb]
\begin{minipage}{.49\linewidth}
    \centering
    \begin{scriptsize}
    \begin{tabularx}{\linewidth}{XXXXX}
\toprule
& Width & Length & Height & Support volume\\
Orientation &&&&\\
\midrule
1  & 6 & 2 &  28 & 10\\
2  & 2 & 28 &  6 & 2\\
3  & 6 & 28 &  2 & 0\\
\bottomrule
\end{tabularx}
    \end{scriptsize}
    \caption{Description of model part data in \citet{che2021machine}}
    \label{tab:0}
\end{minipage}\hfill
\begin{minipage}{.49\linewidth}
        \centering
    \begin{scriptsize}
    \begin{tabularx}{\textwidth}{XXXXXXX}
\toprule
& $W_m$ & $L_m$ & $H_m$ & $T_m^{\mathrm{recoat}}$&  $T_m^{\mathrm{scan}}$& $T_m^{\mathrm{setup}}$\\
Type &&&&\\
\midrule
1  & 25.0 & 25.0 & 32.5 & 0.7  & 0.030864 & 2\\
2  & 40.0 & 80.0 & 50.0 & 0.25 & 0.030864 & 1\\
3  & 40.0 & 40.0 & 40.0 & 0.14 & 0.030864 & 1\\
4  & 40.0 & 60.0 & 45.0 & 0.16 & 0.030864 & 1\\
\bottomrule
\end{tabularx}
    \end{scriptsize}
    \caption{Description of machine data from \citet{che2021machine}}
    \label{tab:-2}
\end{minipage}
\end{table}

To adapt the instances of \citet{che2021machine} to our problem setting, we fix the orientation for each part in advance by considering the calculation of the completion times of a batch in constraints \eqref{E8} and \eqref{E9}. Despite minor variations due to different support volumes, the choice of orientation does not affect the volume-dependent aspect of the processing time. However, taller parts directly influence the height of a batch and thus directly influence the processing time; hence, to speed up the printing process, parts should be designed so that they can be printed with the largest area lying on the build plate. We apply the \acrfull{mhu} rule to decide which orientation to take: for example, the application of \acrshort{mhu} would result in orientation 3 for the part in \Autoref{tab:0}. The respective support volume is added to the total volume $v_i$ of each part $i$. 

\subsubsection{Results}
The general results of all solution approaches for each instance are displayed in \autoref{tab:1}. First, we find that all instances of class 1 are optimally solved by all approaches. However, in comparisons of the average computation times $T^{\mathrm{avg}}$, we observe clear advantages of {\BCo} and {\BCts}: on average, both \acrshort{bnc} variants need less than 50\% of the computation time needed for {\Che} to prove optimality. Looking at the results for class 2, we see that the {\Che} can only prove optimality in at least one run of four instances; in contrast, {\BCo} and {\BCts} provide proven optimal solutions in at least one run for five and six instances, respectively. In general, both \acrshort{bnc} variants yield considerably better lower and upper bounds in the best run than the model {\Che}. However, for {\BCo}, we observe a large deviation between the best and the average optimality gaps for some instances (e.g., 2, 9, and especially 15). In contrast, {\BCts} provides significantly smaller gap values on average and smaller deviations between the best and average values.

\afterpage{
\clearpage
\begin{landscape}
\renewcommand{\arraystretch}{1.05}
\begin{table}[!htbp]
\setlength{\tabcolsep}{3pt}
    \centering
    \begin{scriptsize}
    \begin{threeparttable}
    \begin{tabular}{lrrrrrrrrrrrrrrrrrrrrrrrrr}
\toprule
    & & \multicolumn{8}{c}{{\Che}} & \multicolumn{8}{c}{{\BCo}} & \multicolumn{8}{c}{{\BCts}}  \\
     \cmidrule(lr{0.5em}){3-10} 
     \cmidrule(lr{0.5em}){11-18}
     \cmidrule(lr{0.5em}){19-26}
    Class & \multicolumn{1}{c}{No.} & {\#}o & $U^{\mathrm{best}}$ & $U^{\mathrm{avg}}$ & $L^{\mathrm{best}}$ &  $L^{\mathrm{avg}}$ & $G^{\mathrm{best}}$  & $G^{\mathrm{avg}}$ & $T^{\mathrm{avg}}$\tnote{1} & {\#}o & $U^{\mathrm{best}}$ & $U^{\mathrm{avg}}$ & $L^{\mathrm{best}}$ &  $L^{\mathrm{avg}}$ & $G^{\mathrm{best}}$  & $G^{\mathrm{avg}}$ & $T^{\mathrm{avg}}$\tnote{1} & {\#}o & $U^{\mathrm{best}}$ & $U^{\mathrm{avg}}$ & $L^{\mathrm{best}}$ &  $L^{\mathrm{avg}}$ & $G^{\mathrm{best}}$  & $G^{\mathrm{avg}}$ & $T^{\mathrm{avg}}$\tnote{1}\\
\midrule
1&1&3&20.42&20.42&20.42&20.42&0.00&0.00&7.49&3&20.42&20.42&20.42&20.42&0.00&0.00&2.79&3&20.42&20.42&20.42&20.42&0.00&0.00&2.61\\
1&2&3&21.02&21.02&21.02&21.02&0.00&0.00&6.17&3&21.02&21.02&21.02&21.02&0.00&0.00&3.22&3&21.02&21.02&21.02&21.02&0.00&0.00&2.84\\
1&3&3&30.12&30.12&30.12&30.12&0.00&0.00&6.66&3&30.12&30.12&30.12&30.12&0.00&0.00&4.08&3&30.12&30.12&30.12&30.12&0.00&0.00&2.32\\
1&4&3&41.85&41.85&41.85&41.85&0.00&0.00&7.54&3&41.85&41.85&41.85&41.85&0.00&0.00&2.96&3&41.85&41.85&41.85&41.85&0.00&0.00&2.16\\
1&5&3&25.77&25.77&25.77&25.77&0.00&0.00&6.29&3&25.77&25.77&25.77&25.77&0.00&0.00&2.20&3&25.77&25.77&25.77&25.77&0.00&0.00&2.28\\
1&6&3&29.93&29.93&29.93&29.93&0.00&0.00&1.60&3&29.93&29.93&29.93&29.93&0.00&0.00&1.07&3&29.93&29.93&29.93&29.93&0.00&0.00&1.58\\
1&7&3&34.75&34.75&34.75&34.75&0.00&0.00&9.05&3&34.75&34.75&34.75&34.75&0.00&0.00&3.48&3&34.75&34.75&34.75&34.75&0.00&0.00&2.72\\
1&8&3&36.12&36.12&36.12&36.12&0.00&0.00&6.23&3&36.12&36.12&36.12&36.12&0.00&0.00&1.70&3&36.12&36.12&36.12&36.12&0.00&0.00&3.61\\
1&9&3&29.27&29.27&29.27&29.27&0.00&0.00&5.12&3&29.27&29.27&29.27&29.27&0.00&0.00&2.05&3&29.27&29.27&29.27&29.27&0.00&0.00&2.08\\
1&10&3&41.34&41.34&41.34&41.34&0.00&0.00&0.72&3&41.34&41.34&41.34&41.34&0.00&0.00&0.59&3&41.34&41.34&41.34&41.34&0.00&0.00&0.87\\
1&11&3&41.60&41.60&41.60&41.60&0.00&0.00&10.30&3&41.60&41.60&41.60&41.60&0.00&0.00&4.64&3&41.60&41.60&41.60&41.60&0.00&0.00&4.76\\
1&12&3&37.81&37.81&37.81&37.81&0.00&0.00&4.98&3&37.81&37.81&37.81&37.81&0.00&0.00&2.46&3&37.81&37.81&37.81&37.81&0.00&0.00&1.82\\
1&13&3&31.67&31.67&31.67&31.67&0.00&0.00&16.39&3&31.67&31.67&31.67&31.67&0.00&0.00&4.44&3&31.67&31.67&31.67&31.67&0.00&0.00&5.70\\
1&14&3&25.90&25.90&25.90&25.90&0.00&0.00&8.22&3&25.90&25.90&25.90&25.90&0.00&0.00&3.40&3&25.90&25.90&25.90&25.90&0.00&0.00&1.98\\
1&15&3&31.64&31.64&31.64&31.64&0.00&0.00&8.99&3&31.64&31.64&31.64&31.64&0.00&0.00&3.66&3&31.64&31.64&31.64&31.64&0.00&0.00&4.40\\
1&16&3&32.09&32.09&32.09&32.09&0.00&0.00&14.16&3&32.09&32.09&32.09&32.09&0.00&0.00&1.12&3&32.09&32.09&32.09&32.09&0.00&0.00&3.04\\
1&17&3&31.79&31.79&31.79&31.79&0.00&0.00&3.13&3&31.79&31.79&31.79&31.79&0.00&0.00&1.60&3&31.79&31.79&31.79&31.79&0.00&0.00&1.90\\
1&18&3&32.82&32.82&32.82&32.82&0.00&0.00&5.79&3&32.82&32.82&32.82&32.82&0.00&0.00&3.28&3&32.82&32.82&32.82&32.82&0.00&0.00&2.92\\
1&19&3&30.71&30.71&30.71&30.71&0.00&0.00&7.90&3&30.71&30.71&30.71&30.71&0.00&0.00&15.50&3&30.71&30.71&30.71&30.71&0.00&0.00&2.52\\
1&20&3&32.46&32.46&32.46&32.46&0.00&0.00&8.15&3&32.46&32.46&32.46&32.46&0.00&0.00&3.28&3&32.46&32.46&32.46&32.46&0.00&0.00&2.50\\
\midrule
1&  $Avg.$ &60&31.95&31.95&31.95&31.95&0.00&0.00&7.25&60&31.95&31.95&31.95&31.95&0.00&0.00&3.38&60&31.95&31.95&31.95&31.95&0.00&0.00&2.73 \\
\midrule
2&1&3&79.00&79.00&79.00&79.00&0.00&0.00&941.08&3&79.00&79.00&79.00&79.00&0.00&0.00&28.37&3&79.00&79.00&79.00&79.00&0.00&0.00&19.13\\
2&2&0&74.24&74.39&70.32&70.32&5.28&5.48&t.l.&0&73.03&78.22&73.03&71.07&0.00&8.82&t.l.&0&73.03&73.03&73.03&73.03&0.00&0.00&t.l.\tnote{2}\\
2&3&0&88.03&88.80&85.62&85.62&2.74&3.58&t.l.&0&87.56&90.49&86.39&85.76&1.33&5.15&t.l.&0&87.55&87.55&86.19&86.13&1.55&1.62&t.l.\\
2&4&0&73.28&73.40&70.57&70.55&3.70&3.88&t.l.&0&73.37&74.98&71.18&70.76&2.98&5.56&t.l.&0&72.93&72.95&71.14&70.64&2.46&3.17&t.l.\\
2&5&3&86.63&86.63&86.63&86.63&0.00&0.00&3089.28&2&86.63&89.90&86.63&86.38&0.00&3.65&2197.21&3&86.63&86.63&86.63&86.63&0.00&0.00&39.75\\
2&6&0&65.42&65.42&65.42&65.42&0.00&0.00&t.l.\tnote{2}&0&65.42&69.56&65.42&65.25&0.00&5.53&t.l.&0&65.42&65.42&65.42&65.42&0.00&0.00&t.l.\tnote{2}\\
2&7&3&63.21&63.21&63.21&63.21&0.00&0.00&2547.69&3&63.21&63.21&63.21&63.21&0.00&0.00&62.13&3&63.21&63.21&63.21&63.21&0.00&0.00&85.99\\
2&8&0&81.25&81.45&76.59&76.57&5.73&6.00&t.l.&0&81.88&84.01&77.49&76.82&5.36&8.51&t.l.&3&79.02&79.02&79.02&79.02&0.00&0.00&212.03\\
2&9&0&84.20&84.21&79.86&79.86&5.16&5.16&t.l.&0&82.24&87.54&80.70&80.18&1.87&8.07&t.l.&0&82.58&82.62&80.72&80.49&2.25&2.58&t.l.\\
2&10&0&84.17&84.43&80.96&80.95&3.82&4.12&t.l.&0&83.50&86.36&81.86&81.45&1.96&5.61&t.l.&0&83.30&83.34&81.69&81.65&1.94&2.03&t.l.\\
2&11&0&80.29&80.86&76.78&76.78&4.37&5.04&t.l.&0&79.89&81.48&77.54&77.14&2.94&5.30&t.l.&0&78.78&78.78&76.89&76.83&2.40&2.48&t.l.\\
2&12&0&65.92&66.06&63.40&63.37&3.92&4.07&t.l.&0&65.47&66.75&63.78&63.59&2.58&4.70&t.l.&0&65.74&65.74&63.75&63.03&3.02&4.13&t.l.\\
2&13&0&83.07&83.37&79.96&79.96&3.74&4.08&t.l.&0&82.15&82.69&80.45&80.36&2.07&2.81&t.l.&3&82.11&82.11&82.11&82.11&0.00&0.00&148.47\\
2&14&0&75.73&75.83&72.49&72.46&4.28&4.45&t.l.&0&75.10&75.51&73.09&73.08&2.69&3.22&t.l.&0&75.01&75.04&73.15&73.11&2.49&2.56&t.l.\\
2&15&0&57.71&58.06&55.90&55.90&3.13&3.71&t.l.&0&58.83&65.41&56.21&55.52&4.45&14.46&t.l.&0&57.54&57.58&56.54&56.11&1.74&2.56&t.l.\\
2&16&3&58.16&58.16&58.16&58.15&0.00&0.00&1700.24&3&58.16&58.16&58.16&58.16&0.00&0.00&45.29&3&58.16&58.16&58.16&58.16&0.00&0.00&49.78\\
2&17&0&81.89&81.96&81.10&81.07&1.00&1.08&t.l.&0&81.76&83.85&81.03&80.97&0.90&3.32&t.l.&0&81.76&81.76&81.76&81.53&0.00&0.29&t.l.\\
2&18&0&56.69&56.92&54.99&54.99&3.00&3.40&t.l.&0&56.62&57.05&55.46&55.26&2.05&3.12&t.l.&0&56.24&56.44&55.40&55.31&1.64&2.00&t.l.\\
2&19&0&75.76&75.84&75.07&75.07&0.91&1.01&t.l.&0&75.86&79.58&75.03&74.88&1.09&5.65&t.l.&0&75.71&75.74&75.04&75.02&0.94&0.95&t.l.\\
2&20&0&67.68&68.31&65.48&65.48&3.24&4.14&t.l.&1&67.13&67.46&67.13&66.72&0.00&1.09&3179.63&0&67.13&67.16&67.01&66.64&0.25&0.78&t.l.\\
\midrule
2&  $Avg.$ &12&74.12&74.31&72.08&72.07&2.70&2.96&3294.42&12&73.84&76.06&72.64&72.28&1.61&4.73&2976.59&18&73.54&73.56&72.79&72.65&1.03&1.26&2549.24 \\
\bottomrule
\end{tabular}
    \begin{tablenotes}
    \item {\#}o: Number of optimal solved runs, $U^{\mathrm{best}}$: Best objective value from all runs, $U^{\mathrm{avg}}$: Avg. objective value from all runs, $L^{\mathrm{best}}$: Best lower bound from all runs, $L^{\mathrm{avg}}$: Avg. lower bound from all runs,  $G^{\mathrm{best}}$: Best optimality gap  from all runs (in \%), $G^{\mathrm{avg}}$: Avg. optimality gap  from all runs (in \%), $T^{\mathrm{avg}}$: Avg. computation time from all runs in seconds.
    \item[1] We report t.l. in this column if all runs reached the time limit of 3600 seconds.
    \item[2] In this instance the results of \acrshort{mip} and the {\BCts} are not reported as optimal, because the MIP gap is not below the defined optimality gap of $7\cdot 10^{-6}$.
    \end{tablenotes}
    \caption{Summarized results for Class 1,2 instances of \citet{che2021machine} based on \acrshort{mhu} rule}
    \label{tab:1}
    \end{threeparttable}
    \end{scriptsize}
\end{table}
\end{landscape}

\renewcommand{\arraystretch}{1.05}
\begin{table}[!htbp]
    \centering
    \begin{scriptsize}
    \begin{threeparttable}
    \begin{tabular}{lrrrrrrrrrrrrrr}
\toprule
& \multicolumn{7}{c}{{\BCo}} & \multicolumn{7}{c}{{\BCts}}  \\
     \cmidrule(lr{0.5em}){2-8} 
     \cmidrule(lr{0.5em}){9-15}
\multicolumn{1}{c}{No.} & {\#}E & \multicolumn{1}{c}{$T_{\mathrm{LB}}$} & $T_{\mathrm{OR}}$ & $T_{\mathrm{BR}}$ & $T_{\mathrm{OP}}$ & $t^{\mathrm{max}}_{\mathrm{OP}}$ & {\#}HP & {\#}E & \multicolumn{1}{c}{$T_{\mathrm{LB}}$} & $T_{\mathrm{OR}}$ & $T_{\mathrm{BR}}$ & $T_{\mathrm{OP}}$ & $t^{\mathrm{max}}_{\mathrm{OP}}$ & {\#}HP \\
\midrule
1&18.33&0.19&1.59&11.91&8.81&1.81&0.00&4.00&0.02&0.24&1.10&1.57&1.47&0.00\\
2&33.67&0.20&0.90&12.98&2976.67&2916.53&1.00&23.67&0.12&0.59&5.26&1168.33&1142.50&0.33\\
3&19.33&0.13&0.57&9.62&3580.42&3548.95&1.00&9.33&0.08&0.26&2.92&3572.70&3530.21&1.00\\
4&28.00&0.08&0.80&7.10&3583.47&3562.15&1.00&17.00&0.09&0.53&9.19&3538.37&3359.60&1.33\\
5&17.33&0.21&0.65&20.94&2170.35&2163.00&0.67&4.00&0.03&0.15&4.75&2.64&2.50&0.00\\
6&18.00&0.22&0.87&13.81&1223.32&1209.46&0.33&4.00&0.03&0.13&1.95&1.46&0.84&0.00\\
7&15.67&0.21&0.74&16.24&6.06&1.14&0.00&4.67&0.05&0.19&4.65&1.26&0.89&0.00\\
8&26.00&0.12&0.49&8.79&3577.22&1716.33&2.33&22.33&0.07&0.65&6.19&114.75&98.76&0.00\\
9&21.67&0.09&0.73&9.57&3585.00&3576.43&1.00&11.67&0.04&0.39&6.32&3547.91&3341.29&1.33\\
10&32.67&0.19&0.76&8.41&3582.20&3169.94&1.33&14.00&0.09&0.38&5.26&3551.57&3435.25&1.00\\
11&29.00&0.13&1.12&9.98&3572.14&3068.59&1.33&10.00&0.03&0.37&1.85&3574.18&3464.07&1.00\\
12&28.67&0.16&0.96&10.26&3568.76&2460.56&2.33&9.67&0.03&0.31&3.73&3556.50&3379.45&1.33\\
13&22.67&0.10&0.59&5.92&3579.56&3535.79&1.00&15.67&0.09&0.48&1.90&34.99&34.03&0.00\\
14&32.00&0.14&1.04&9.37&3584.15&3521.32&1.00&16.33&0.11&0.48&5.58&3390.62&3181.37&1.33\\
15&12.00&0.08&0.28&3.55&3594.60&3158.34&1.33&10.67&0.08&0.31&3.54&3564.32&3535.48&1.00\\
16&20.00&0.30&1.23&24.71&6.85&1.10&0.00&4.00&0.04&0.16&3.67&0.93&0.56&0.00\\
17&20.00&0.19&0.58&4.06&3592.15&3281.80&1.33&10.33&0.06&0.26&1.22&1077.48&1075.05&0.33\\
18&19.00&0.14&1.08&10.65&3582.76&3018.78&1.33&12.67&0.11&0.66&3.61&3475.80&3179.99&1.33\\
19&12.67&0.15&0.38&9.61&3589.09&3554.51&1.00&9.00&0.12&0.27&9.94&3229.04&3086.27&1.00\\
20&30.33&0.21&1.48&21.41&3130.76&1971.02&2.33&12.00&0.09&0.60&5.45&3541.27&3379.32&1.33\\
\bottomrule
\end{tabular}
\begin{tablenotes}
    \item  {\#}E: Number of batches checked. $T_{\mathrm{LB}}$: Total computation time spent on lower bounds, $T_{\mathrm{OR}}$: Total computation time spent on orthogonal relaxation, $T_{\mathrm{BR}}$: Total computation time spent on bar relaxation, $T_{\mathrm{OP}}$: Total computation time spent on orthogonal packing, $t^{\mathrm{max}}_{\mathrm{OP}}$: Maximum computation time spent on orthogonal packing, {\#}HP: Number of hard packing instances with $T_{\mathrm{OP}} > 300$.
    \end{tablenotes}
\caption{Detailed information on computation times for Class 2 instances}
    \label{tab:detailed}
    \end{threeparttable}
    \end{scriptsize}
\end{table}

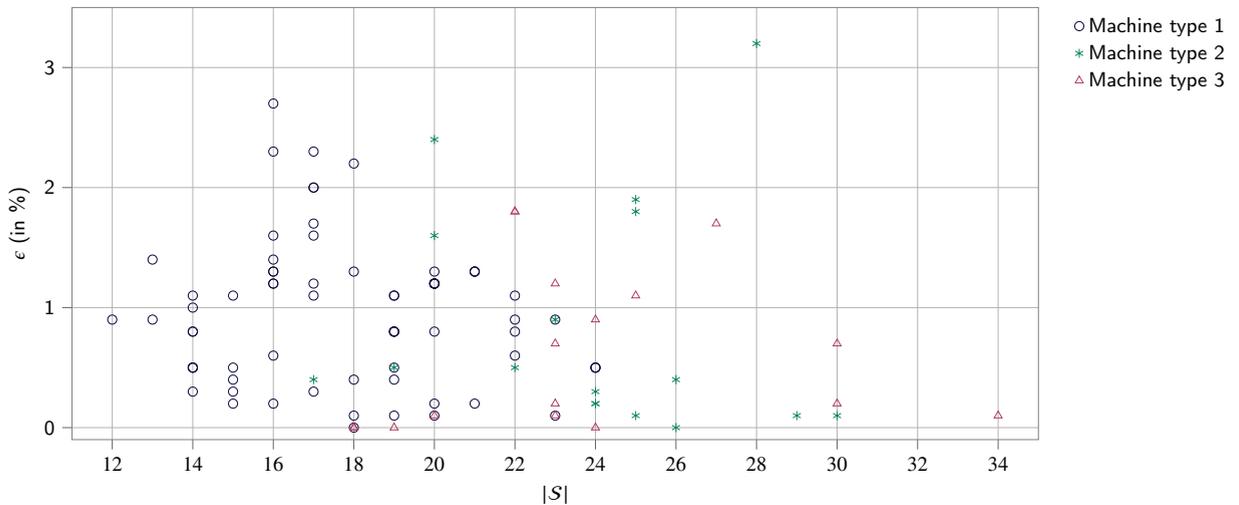
\begin{figure}[!htb]
         \centering
         \begin{tikzpicture}
           \node[right = 5cm of 2] (3){\resizebox{\textwidth}{!}{
\pgfplotsset{scaled y ticks=false}
\begin{tikzpicture}
\begin{axis}[
tick align=outside,
tick pos=left,
label style = {text =black},
tick label style = {text =black},
xlabel={$|\mathcal{S}|$},
ylabel={$\epsilon $ (in \%)},
x grid style={gray2, thin},
xmin=11, xmax=35,
xtick style={color=gray1, text =black},
y grid style={gray2, thin},
ymin=-0.001, ymax=0.035,
ytick = {0, 0.01, 0.02, 0.03},
yticklabels= {0, 1, 2, 3},
ytick style={color=gray1, text =black},
legend cell align=left,
legend pos=outer north east,
legend style={draw=none, text =black},
ymajorgrids=true,
xmajorgrids=true,
width=\textwidth,
height=0.5\textwidth,
]
\addplot [draw=color2, fill=color2, mark=o, only marks]
table{%
x  y
13 0.00900000000000001
14 0.011
14 0.00800000000000001
14 0.00800000000000001
15 0.005
16 0.00600000000000001
17 0.003
19 0.00800000000000001
19 0.00800000000000001
19 0.00800000000000001
16 0.012
17 0.011
15 0.003
19 0.011
18 0.001
18 0
12 0.00900000000000001
20 0.001
19 0.001
20 0.002
22 0.00600000000000001
22 0.00800000000000001
24 0.005
24 0.005
14 0.005
20 0.012
14 0.005
14 0.01
18 0.013
14 0.005
20 0.012
16 0.012
19 0.00800000000000001
21 0.013
16 0.013
16 0.016
16 0.023
15 0.004
15 0.011
20 0.013
14 0.003
16 0.014
18 0.004
13 0.014
16 0.027
17 0.02
17 0.02
19 0.004
17 0.017
17 0.016
22 0.011
22 0.00900000000000001
21 0.002
21 0.013
24 0.005
23 0.001
23 0.00900000000000001
20 0.012
20 0.012
20 0.012
20 0.012
19 0.005
19 0.011
20 0.00800000000000001
21 0.013
15 0.002
17 0.023
16 0.013
17 0.012
16 0.002
18 0.022
};
\addplot [draw=color1, fill=color1, mark=asterisk, only marks]
table{%
x  y
28 0.032
24 0.003
25 0.018
26 0.004
20 0.024
30 0.001
23 0.00900000000000001
25 0.001
17 0.004
29 0.001
24 0.002
25 0.019
26 0
24 0.002
20 0.016
19 0.005
22 0.005
};
\addplot [draw=color0, fill=color0, mark=triangle, only marks]
table{%
x  y
23 0.001
24 0
19 0
30 0.00700000000000001
34 0.001
27 0.017
18 0
22 0.018
23 0.002
18 0
18 0
24 0.00900000000000001
25 0.011
23 0.012
30 0.002
20 0.001
22 0.018
23 0.00700000000000001
};
\addplot [draw=black, fill=black, mark=*, only marks]
table{%
x  y
};
\legend{Machine type 1, Machine type 2, Machine type 3, Machine type 4}
\end{axis}

\end{tikzpicture}}};
         \end{tikzpicture}
        \caption{Scatter plot representing $\epsilon$ values and item numbers of hard packing instances ($T_{OP} > 300$ sec.) for the \acrshort{bnc} variants}
        \label{fig:EpsilonScatter}
\end{figure} 

}

In what follows, we examine the results in more detail to determine the reasons for these performance differences between the \acrshort{bnc} variants. To this end, \Autoref{tab:detailed} provides information about the times spent in the different feasibility checks of the algorithms. We use the following notations: \#E represents the number of explored batch assignments; $T_{\mathrm{LB}}$,  $T_{\mathrm{OR}}$, $T_{\mathrm{BR}}$, and $T_{\mathrm{OP}}$ represent the total computation times of the lower bound (LB), orthogonal relaxation (OR), bar relaxation (BR), and orthogonal packing (OP) verification methods; $t^{\mathrm{max}}_{\mathrm{OP}}$ denotes the maximum time spent for a single orthogonal packing problem averaged by all seed runs; and {\#}HP reports the number of hard packing instances. We define a hard packing instance as an orthogonal packing problem that takes at least 300 sec to be solved in the \acrshort{cp} model. All numbers represent average values over three seed runs. It should be noted that we can only terminate the algorithm after checking an integer solution; therefore, total times of slightly more than 3,600 sec may also be reported.

First, we observe that in both approaches, only a small computational effort is used for the relaxed orthogonal packing checks LB, OR, and BR. In most instances, the main share of the computation time is spent in the exact \acrshort{2dopr} check: column $t^{\mathrm{max}}_{\mathrm{OP}}$ shows that in several problem instances, the available computation time is nearly entirely consumed by a single check of the \acrshort{2dopr}. However, in some instances (e.g., 2, 5, 6, 8, 13) the variant {\BCts} spends significantly less time in the exact \acrshort{2dopr} check, and $t^{\mathrm{max}}_{\mathrm{OP}}$ is greatly reduced. Additionally, we observe that the number of hard packing instances is slightly reduced in the variant {\BCts}. Thus, it can be deduced that the original variant {\BCo} gets stuck more often in the exact check of batches very early in the solution process; as a result, the lower and upper bounds cannot improve as much as they do in the variant {\BCts}. We also can see that hard packing instances are not necessarily part of an optimal solution. For example, instances 8 and 13 can be solved to optimality with {\BCts} in fast computation times, while {\BCo} spends much time in the exact \acrshort{2dopr} check. Providing a better initial solution can thus be beneficial, since it helps to prune the search tree.

We further investigate the hard packing instances to identify possible reasons for why the \acrshort{cp} solver cannot decide whether a \acrshort{2dopr} instance is feasible or infeasible in a reasonable time. Following \citet{CLAUTIAUX20071196}, we calculate measure $\epsilon$ for each packing problem with the set of parts $\mathcal{S}$ on a machine $m$ that has a runtime of at least 300 sec, where $\sum_{i \in \mathcal{S}} a_i = (1 - \epsilon) \cdot A_m$, meaning that $\epsilon$ measures the free area when considering a specific batch. \Autoref{fig:EpsilonScatter} shows the aggregated results as scatter plots, with measures $\epsilon$ and $|\mathcal{S}|$ segmented by the different machine types. All hard-to-solve packing problems except one have $\epsilon$ values below 3\%, implying a high packing density between 97 -- 100\% in the considered batches. These results are consistent with the findings of \citet{CLAUTIAUX20071196}, but difficult instances are slightly skewed toward lower $\epsilon$. The authors show in their computational experiments that a high area utilization is associated with increased computational difficulty and therefore impacts whether a packing problem can be solved in a reasonable time. Examining the different machine types, we observe that most hard packing problems occur on machine type 1 with the smallest build platform, while fewer hard packing instances are seen with increasing production areas of machine types. For the machine type 4, the largest in terms of the build area, no hard packing instance is detected in the experiments. Additionally, the number of parts in the hard packing problems indicates that the parts are small compared to the machine area. To validate this assumption, we give a more detailed overview of the instances introduced by \citet{che2021machine} in the following section. We also analyze other instance sets used for scheduling in additive manufacturing in order to give a broader perspective.

\subsection{Generation of new benchmark instances}
\subsubsection{Discussion on instance data used for scheduling in additive manufacturing}
We examine the data from various studies on \acrshort{am} scheduling to gain insights on the part dimensions they use. \autoref{tab:6} shows selected problem data for planning in \acrshort{am} that is either accessible online via provided links or explicitly explained in the studies. The first column presents the source of the problem data, and column $\#Parts$ describes the total number of parts evaluated. The next five columns present the evaluation of part areas: this begins with the average part area $\bar{\alpha_i}$, followed by percentiles and the maximum part area of the considered data displayed in columns $P_{25\%}$, $P_{50\%}$, $P_{75\%}$, and $\max_{i \in I} \alpha_i$. The remaining columns of the table report shares relative to the minimum machine area of each problem instance: the first column in this section presents the average share of area taken up by a part, while the last two columns indicate the share of parts that are smaller than 50\% and 25\% of the minimum build area, respectively.

\Autoref{tab:6} reports the smallest part areas for the data of \citet{Chergui2018} and \citet{che2021machine}. This applies for both the absolute part areas and the relative areas with regard to the minimum machine area. Comparing the interquartile ranges of those data sets to the ones from \citet{Kucukkoc2019}, we can further deduce less diversity in part sizes in the data of \citet{Chergui2018} and \citet{che2021machine}. Evaluating the relative part areas to the minimum machine area, we see that the majority of parts occupy less than 25\% of the minimum machine area. 

Based on these findings, we see two drawbacks of the used test set. First, a test set should comprise a diverse set of instances in order to study, e.g., the influences of different part sizes on the solution process and the solution quality. A more diversified benchmark data set that takes into account varying part dimensions may also more precisely represent real-world requirements in \acrshort{am} shops. Second, test instances with a majority of small parts might make solving for the proposed \acrshort{bnc} more difficult, because many techniques for preprocessing, lower bounds determination, and domain reduction leverage parts that are larger than 50\% of the bin dimensions. Thus, these techniques have no effect on the instances of \citet{che2021machine}; the effectiveness of the proposed \acrshort{bnc} variants cannot be tested properly with the test data, since the approaches get stuck early in the search. As can be seen in \Autoref{tab:6}, all of the remaining available data assumes one-dimensional packing. Consequently, a new benchmark data set is needed for planning in \acrshort{am}, one that incorporates different part sizes in accordance with the well-known benchmark data for cutting and packing problems \citep[see, e.g.,][]{berkey1987two,martello1998exact}. We propose such a new benchmark data set in the next section.
\begin{table}[!htb]
    \centering
    \begin{scriptsize}
    \begin{threeparttable}
    \begin{tabularx}{\textwidth}{llrrrrrrrPPP}
\toprule
&&&& \multicolumn{5}{c}{Part areas $\alpha_i$ (in $cm^{2}$)} & \multicolumn{3}{c}{Part area shares (in \%)}\\
     \cmidrule(lr{0.5em}){5-9}
     \cmidrule(lr{0.5em}){10-12}
Reference & data set & {\#}D & {\#}Parts& $\bar{\alpha_i}$&$P_{25\%}$&$P_{50\%}$&$P_{75\%}$ & $\max \alpha_i$ & $\frac{\bar{\alpha_i}}{\min\{A_m\}}$&$S_{50\%}$ & $S_{25\%}$ \\
\midrule
\citet{rohaninejad2021scheduling} & -     & 1 &1985 &  500.64 & 399.24 &  500.08 &  601.04 &  966.44 &  59.13 &   31.34 &   2.27 \\
\citet{che2021machine}            & $ht1$ & 2 & 400 &   73.10 &  25.00 &   38.22 &   88.00 &  289.80 &  10.14 &  100.00 &  86.25 \\
\citet{che2021machine}            & $ht2$ & 2 &1000 &   65.46 &  25.00 &   38.00 &   68.89 &  289.80 &   7.81 &  100.00 &  91.70 \\
\citet{Kucukkoc2019}              & $SM$  & 1 & 126 &  213.45 &  91.35 &  248.92 &  269.75 &  709.06 &  28.79 &   89.68 &  42.86 \\
\citet{Kucukkoc2019}              & $IM$  & 1 & 638 &  227.61 &  89.68 &  174.43 &  270.44 &  980.08 &  20.59 &   88.09 &  75.86 \\
\citet{Kucukkoc2019}              & $UM$  & 1 & 638 &  227.61 &  89.68 &  174.43 &  270.44 &  980.08 &  28.16 &   85.42 &  60.03 \\
\citet{Chergui2018}               & -     & 2 &  20 &   67.51 &  15.08 &   35.65 &   51.88 &  361.00 &  10.80 &   90.00 &  90.00 \\
\bottomrule
\end{tabularx}
    \begin{tablenotes}
    \item  {\#}D: Number of dimensions in packing, {\#}Parts: Number of evaluated part specifications, $\bar{a_i}$: Avg. part area, $P_{25\%}$: 25\% percentile of part area within the test data, $P_{50\%}$: 50\% percentile of part area within the test data, $P_{75\%}$: 75\% percentile of part area within the test data, $\max a_i$: maximum part area within the test data, $\frac{\bar{a_i}}{\min\{A_m\}}$: Avg. area consumption of one part based on the minimum machine area, $S_{50\%}$: Share of parts that are smaller than 50\% of the minimum machine area in the test data, $S_{25\%}$: Share of parts that are smaller than 25\% of the minimum machine area in the test data.
    \end{tablenotes}
    \caption{Comparison of part data of available test data in literature on scheduling in additive manufacturing}
    \label{tab:6}
    \end{threeparttable}
    \end{scriptsize}
\end{table}


\subsubsection{Description of new test instances}
We first define the machine system and the part sizes for the new test data. The structure of the available machines is explained in \Autoref{tab:machineGenerationDetails} in the appendix. 
As regards the dimensions of the machines' build spaces, we follow the product data sheets of machine manufacturers \citep[see, e.g.,][]{SLMSolutionsAG} and the machine specifications from related work \citep{Kucukkoc2019, che2021machine}. We randomly draw machine parameters from the intervals shown in the table to generate the process speed parameters; these intervals are also based on manufacturers' product data sheets. We then use the formulas proposed by \citet{aloui2021heuristic} to calculate scanning and recoating times.

As regards part sizes, we follow the approach of \citet{martello1998exact}, who define different part types and create classes in which certain types of parts occur more frequently. This allows us to generate a diverse set of parts. Furthermore, we can examine the solution behavior of the algorithms based on different part sizes. \Autoref{tab:partGenerationDetails} gives a detailed explanation of the four part types and the classes of the test data. According to the definition, classes 1 and 2 contain rather small parts, while classes 3 and 4 comprise bigger parts, and the parts in classes 2 and 3 tend to be more elongated compared to those in classes 1 and 4. Finally, we define the test instances in \Autoref{tab:instanceCreation}, resulting in 36 configurations and a total of 180 problem instances. For each instance, we ensure that the build space of machine type 1 is included for at least one machine; the remaining machine build spaces are drawn randomly from the available types in \Autoref{tab:machineGenerationDetails}. In this way, we ensure that each part fits in at least one machine.\footnote{All test instances are available online via:\\ \url{https://data.mendeley.com/datasets/k4vvbvf5kb/draft?a=07c17363-17e5-424f-9aa9-d59328005ae0}}

\renewcommand{\arraystretch}{1.05}
\begin{table}[!htb]
\begin{footnotesize}
\begin{tabularx}{\textwidth}{lR}
    \toprule
		Part type 1 & uniformly random integers: $\omega_{i}$ in [1, $\frac{1}{5}\Omega$],  $\lambda_{i}$ in [1, $\frac{1}{5}\Lambda$], $h_i$ in [1, $\frac{1}{2}H$]\\
		Part type 2 & uniformly random integers: $\omega_{i}$ in [1, $\frac{1}{5}\Omega$],  $\lambda_{i}$ in [$\frac{1}{5}\Lambda$, $\frac{3}{5}\Lambda$], $h_i$ in [1, $\frac{1}{2}H$]\\
		Part type 3 & uniformly random integers: $\omega_{i}$ in [$\frac{1}{5}\Omega$, $\frac{2}{5}\Omega$],  $\lambda_{i}$ in [$\frac{1}{5}\Lambda$, $\frac{4}{5}\Lambda$], $h_i$ in [1, $\frac{1}{2}H$]\\
		Part type 4 & uniformly random integers: $\omega_{i}$ in [$\frac{1}{5}\Omega$, $\frac{3}{5}\Omega$],  $\lambda_{i}$ in [$\frac{1}{5}\Lambda$, $\frac{3}{5}\Lambda$], $h_i$ in [1, $\frac{1}{2}H$]\\
		\midrule
		Class 1 & part type 1 with probability 70\%, type 2, 3, 4 with probability 10\% each. $\Omega = \Lambda = H = 50$\\
		Class 2 & part type 2 with probability 70\%, type 1, 3, 4 with probability 10\% each. $\Omega = \Lambda = H = 50$\\
		Class 3 & part type 3 with probability 70\%, type 1, 2, 4 with probability 10\% each. $\Omega = \Lambda = H = 50$\\
		Class 4 & part type 4 with probability 70\%, type 1, 2, 3 with probability 10\% each. $\Omega = \Lambda = H = 500$\\
        \bottomrule
	\end{tabularx}
	\end{footnotesize}
	\caption{Definition of part types and classes}
	\label{tab:partGenerationDetails}
\end{table}
\begin{table}[!htb]
\begin{footnotesize}
\begin{tabularx}{\textwidth}{Rlr}
    \toprule
    \multicolumn{3}{l}{\textbf{Parameter and levels for test data generation}}\\
        \midrule
		Parameters & Levels & Total\\
		\midrule 
		&&\\
		Combinations of parts and machines & {\begin{tabularx}{8cm}{RXXXXX} 
		\toprule
		& \multicolumn{5}{c}{$\vert \mathcal{I} \vert$}\\
		\cmidrule{2-6}
		$\vert \mathcal{M} \vert$ & 10 & 20 & 40 & 60 & 80 \\
		\midrule
		2 & X & X & X & X & X\\
		3 &   & X & X &   &  \\
		5 &   &   &   & X & X\\
		\bottomrule
		\end{tabularx}}& 9 \\
		&&\\
		Number of part classes & 1, 2, 3, 4 & 4 \\
	    \midrule
	    & Total number of configurations & 36\\
	    & Number of instances/ configuration & 5 \\
	    & Number of problem instances & 180 \\
	    & Number of runs/ instance & 3 \\
	    & Total number of experiments & 540 \\
        \bottomrule
	\end{tabularx}
	\end{footnotesize}
	\caption{Parameters and their levels for test data and parameter tuning}
	\label{tab:instanceCreation}
\end{table}

\subsection{Results for new benchmark instances}
\begin{figure}[htb]
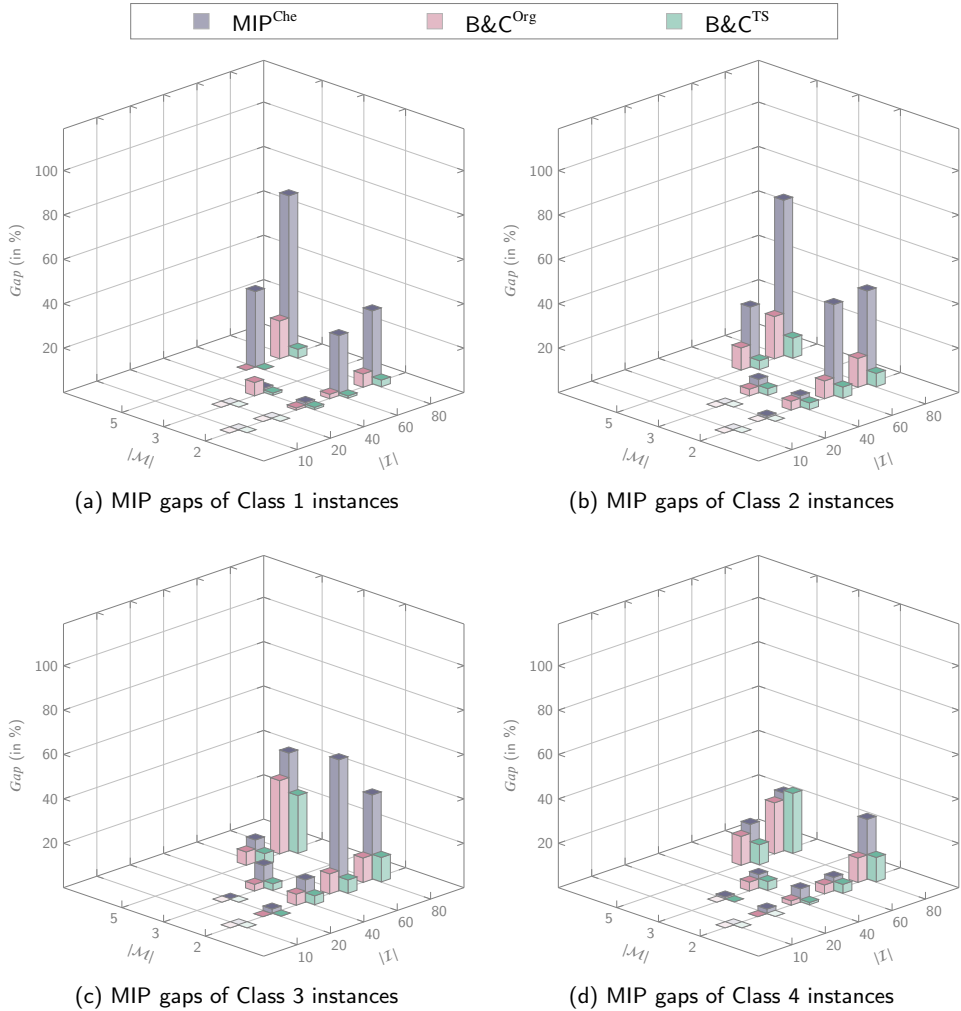

    \centering
    \begin{tikzpicture}
     \node[](1){\resizebox{0.8\textwidth}{!}{\begin{tikzpicture}
\node [] (1) at (0.0, 7.0) {\begin{tikzpicture}[node distance = 3cm]
\node[](1){\resizebox{0.4\textwidth}{!}{\input{Class1_BarPlot}}};
\end{tikzpicture}};
\node [] (2) at (7.0, 7.0) {\begin{tikzpicture}[node distance = 3cm]
\node[](1){\resizebox{0.4\textwidth}{!}{\input{Class2_BarPlot}}};
\end{tikzpicture}};
\node [black] (3) at (0.0, 3.7) {\sffamily (a) MIP gaps of Class 1 instances};
\node [black] (4) at (7.0, 3.7) {\sffamily (b) MIP gaps of Class 2 instances};
\node [] (5) at (0.0, 0.0) {\begin{tikzpicture}[node distance = 3cm]
\node[](1){\resizebox{0.4\textwidth}{!}{\input{Class3_BarPlot}}};
\end{tikzpicture}};
\node [] (6) at (7.0, 0.0) {\begin{tikzpicture}[node distance = 3cm]
\node[](1){\resizebox{0.4\textwidth}{!}{\input{Class4_BarPlot}}};
\end{tikzpicture}};
\node [black] (7) at (0.0, -3.3) {\sffamily (c) MIP gaps of Class 3 instances};
\node [black] (8) at (7.0, -3.3) {\sffamily (d) MIP gaps of Class 4 instances};
\node [] (Legend) at (3.5, 10.5) {\begin{tikzpicture}[scale=1]

\begin{scope}
\node[draw, rectangle, minimum width = 10cm, minimum height = 0.5cm] (0) at (0.0, 0.0) {};
\node[draw, rectangle, fill = color2!60, opacity = 0.6, minimum width = 0.1cm, minimum height = 0.1cm] (1) at (-4.0, 0.0) {};
\node[right = 0.2cm of 1, black] (2) {\Che};
\node[draw, rectangle, right = 1.7cm of 2, fill = color0!60, opacity = 0.6, minimum width = 0.1cm, minimum height = 0.1cm] (3) {};
\node[right = 0.2cm of 3, black] (4) {\BCo};
\node[draw, rectangle, right = 1.7cm of 4, fill = color1!60, opacity = 0.6, minimum width = 0.1cm, minimum height = 0.1cm] (5) {};
\node[right = 0.2cm of 5, black] (6) {\BCts};

\end{scope}
\end{tikzpicture}
};
\node [anchor=south] (Boxes-3) at (3) {
};
\end{tikzpicture}}};
    \end{tikzpicture}
\caption{3D bar plots illustrating the resulting optimality gaps by classes, item numbers, and machine numbers}
\label{fig:3dPlotsByClasses}
\end{figure}

\Autoref{fig:3dPlotsByClasses} presents an overview of the test results, reporting the optimality gaps at the termination of each method; the results are segmented by classes, number of parts $\mathcal{I}$, and number of machines $\mathcal{M}$. More detailed results for the different approaches are given in \autoref{tab:newTestsdetailed}.\footnote{A comprehensive summary of all test results can be found in the supplementary materials of this manuscript. The individual output files are available online via: \url{https://data.mendeley.com/datasets/k4vvbvf5kb/draft?a=07c17363-17e5-424f-9aa9-d59328005ae0}}
Overall, we deduce from \Autoref{fig:3dPlotsByClasses} that both \acrshort{bnc} variants generate superior results to the \acrshort{mip} model for almost all configurations. This superiority is especially pronounced for larger instance sizes. As a result, the decomposition is particularly useful for instances with larger problem sizes. For the {\Che} model, we can see a sharp increase in the optimality gaps as the number of parts increases across all classes; in contrast, the optimality gaps of both \acrshort{bnc} variants grow only moderately with the instance size. For classes 1 and 2, both \acrshort{bnc} variants generate better results than {\Che}, but the variant {\BCts} provides significantly smaller MIP gaps than {\BCo}. In these classes, {\Che} produces MIP gaps up to 72\% for the largest instances, whereas  {\BCo} and {\BCts} result in average gaps below 20\% and below 10\%, respectively. For classes 3 and 4, we see that the MIP gaps of both \acrshort{bnc} variants are larger and more uniform than those in classes 1 and 2, while the MIP gaps of {\Che} are mostly smaller than those in classes 1 and 2; thus, the overall benefits of the \acrshort{bnc} variants decrease. However, it is only for the largest problem size of class 4 that no performance advantages of the \acrshort{bnc} variants over {\Che} can be seen.

The reason for these differences is the varying relevance of both problems in additive manufacturing -- i.e., packing and scheduling. Recall that classes 1 and 2 include small parts; thus, more parts are packed into a single batch, and fewer batches are needed to produce all parts. Here, on the one hand, the packing problem is more relevant and harder to solve, while on the other hand, the scheduling problem (i.e., which batch is produced on which machine at which time) is less important, since fewer batches results in fewer possibilities to combine batches. For classes 3 and 4, fewer parts fit into a batch, which means more batches are required. Here, the scheduling decision is crucial, while the packing feasibility check is easier and less relevant. Since the master problem manages the scheduling decisions, the results of the \acrshort{bnc} variants and {\Che} are more similar in classes 3 and 4. We also hypothesize that the obtained lower bounds in classes 3 and 4 are weaker than those in classes 1 and 2 due to the more prominent scheduling problem. In the comparison of the \acrshort{bnc} variants, the better optimality gaps of {\BCts} in classes 1 and 2 are primarily the result of better upper bounds of the final solutions in {\BCts}. These improved upper bounds result from better initial solutions by invoking the more restricted problem first, and better initial solution prevent the algorithm in {\BCts} from checking lower-quality solutions with dense packings. In this way, compared to {\BCo}, {\BCts} is able to check more solutions with better upper bounds.

These findings are supported by the detailed analysis of the added cuts and the time needed to check feasibility  (see \autoref{tab:newTestsCuts}). For classes 1 and 2, a large share of the total runtime is spent in the feasibility check, whereas for classes 3 and 4, this share is much smaller, even though far more batches are checked. We can also observe that fewer batches are explored for instances with 80 parts compared to instances with 60 parts for classes 3 and 4. Solving the continuous relaxation at each node of the branch-and-bound tree becomes computationally more demanding with increasing instance size; as a result, fewer nodes and fewer integer solutions can be explored, which in turn leads to a deterioration of the solution quality for larger instances. 

With regard to the added cuts, the simple lower bounds are ineffective for instances with smaller parts but are useful if the parts are larger. In contrast, the orthogonal relaxation procedure can prove infeasibility for a large share of batches across all classes. Despite the preceding orthogonal relaxation using \acrshort{dffs}, the bar relaxation also generates a substantial number of cuts and prevents the algorithm from unnecessary calls of the \acrshort{2dopr}. Evaluating the different classes, we see that the bar relaxation is more advantageous with larger parts. The improved initial solution in {\BCts} significantly impacts the number of explored batches {\#}E and, consequently, the check time. Except for the instances with 80 parts in classes 3 and 4, {\BCts} demonstrates a significantly lower time spent on feasibility checks than {\BCo}. 

\afterpage{
\clearpage
\begin{landscape}
\renewcommand{\arraystretch}{1.01}
\begin{table}[!htb]
\setlength{\tabcolsep}{3pt}
\newcolumntype{U}{>{\raggedleft\arraybackslash}p{0.33in}}
\newcolumntype{T}{>{\raggedleft\arraybackslash}p{0.3in}}
\newcolumntype{G}{>{\raggedleft\arraybackslash}p{0.18in}}
\newcolumntype{O}{>{\raggedleft\arraybackslash}X}
    \centering
    \begin{scriptsize}
    \begin{threeparttable}

    \begin{tabularx}{1.36\textwidth}{lllrOOOrrOOOOrrOOOOrr}
\toprule
&&&& \multicolumn{5}{c}{{\Che}} & \multicolumn{6}{c}{{\BCo}} & \multicolumn{6}{c}{{\BCts}}\\
\cmidrule(lr{0.5em}){5-9}
\cmidrule(lr{0.5em}){10-15}
\cmidrule(lr{0.5em}){16-21}
Cl. & $\vert \mathcal{I} \vert$ & $\vert \mathcal{M} \vert$ & {\#}runs  & $U^{\mathrm{sum}}$ & $L^{\mathrm{sum}}$ & $G^{\mathrm{avg}}$ & $T^{\mathrm{avg}}$ & {\#}o  & $U_0^{\mathrm{sum}}$ & $U^{\mathrm{sum}}$ & $L^{\mathrm{sum}}$ & $G^{\mathrm{avg}}$ & $T^{\mathrm{avg}}$ & {\#}o   & $U_0^{\mathrm{sum}}$ & $U^{\mathrm{sum}}$ & $L^{\mathrm{sum}}$ & $G^{\mathrm{avg}}$ & $T^{\mathrm{avg}}$ & {\#}o \\
\midrule
1& 10 & 2 & 15 &400.44&400.44&0.00&0.37&15&413.00&400.44&400.44&0.00&0.40&15&400.44&400.44&400.44&0.00&0.52&15\\
 & 20 & 2 & 15 &333.18&333.18&0.00&3.81&15&352.02&333.18&333.18&0.00&1.97&15&333.18&333.18&333.18&0.00&1.70&15\\
 &    & 3 & 15 &371.66&371.66&0.00&7.74&15&497.16&371.66&371.66&0.00&4.51&15&371.80&371.66&371.66&0.00&5.42&15\\
 & 40 & 2 & 15 &819.55&806.27&1.78&2922.79&3&923.65&817.75&813.41&0.94&1950.17&9&820.37&817.93&812.54&0.97&1910.90&8\\
 &    & 3 & 15 &939.25&930.03&1.57&2219.14&6&1216.49&959.48&928.14&6.25&1492.87&9&937.88&936.16&933.50&0.67&1231.02&11\\
 & 60 & 2 & 15 &2748.81&1650.81&27.33&3431.97&3&1958.92&1734.05&1708.13&1.69&2669.16&5&1731.65&1725.53&1712.68&0.84&2884.98&3\\
 &    & 5 & 15 &3914.39&444.26&33.97&2946.50&6&828.93&471.26&471.21&0.01&2185.40&6&471.84&471.26&471.22&0.01&2196.12&6\\
 & 80 & 2 & 15 &3455.56&2136.54&33.44&3600.33&0&2564.01&2341.20&2188.16&5.81&3602.18&0&2266.03&2261.30&2193.57&2.63&3600.78&0\\
 &    & 5 & 15 &8869.13&635.44&72.37&3600.30&0&1139.63&777.34&682.88&10.88&3601.17&0&724.41&718.63&690.79&4.08&3601.16&0\\
\midrule
2& 10 & 2 & 15 &262.80&262.80&0.00&0.38&15&289.28&262.80&262.80&0.00&0.51&15&263.24&262.80&262.80&0.00&0.60&15\\
 & 20 & 2 & 15 &650.18&646.63&0.61&1327.87&10&735.88&648.91&648.91&0.00&34.55&15&655.87&648.91&648.91&0.00&33.54&15\\
 &    & 3 & 15 &486.49&486.49&0.00&55.07&15&624.70&486.49&486.49&0.00&10.35&15&488.27&486.49&486.49&0.00&13.75&15\\
 & 40 & 2 & 15 &1560.58&1490.98&4.90&3349.67&2&1709.00&1551.33&1509.13&3.48&2884.32&3&1631.65&1539.82&1508.01&2.58&2885.24&3\\
 &    & 3 & 15 &932.66&872.40&6.44&3600.74&0&1306.83&913.99&888.14&2.88&3600.87&0&913.42&909.70&889.02&2.47&3401.18&1\\
 & 60 & 2 & 15 &3619.80&1762.20&40.96&3601.29&0&2185.17&1990.27&1828.42&8.08&3601.32&0&1926.58&1925.29&1828.52&4.88&3600.87&0\\
 &    & 5 & 15 &1835.50&877.22&27.21&3512.13&1&1523.55&998.01&903.16&9.55&2194.78&6&944.08&939.41&909.54&3.68&2195.39&6\\
 & 80 & 2 & 15 &5440.44&2658.27&41.93&3600.24&0&3368.21&3184.02&2780.14&13.33&3600.77&0&2953.38&2951.43&2793.33&5.67&3600.86&0\\
 &    & 5 & 15 &11716.41&1003.81&69.98&3600.60&0&1750.67&1335.01&1065.16&18.79&3601.10&0&1175.29&1172.12&1065.98&8.81&3600.65&0\\
\midrule
3& 10 & 2 & 15 &1090.17&1090.17&0.00&4.33&15&1125.32&1090.17&1090.17&0.00&1.05&15&1090.17&1090.17&1090.17&0.00&1.33&15\\
 & 20 & 2 & 15 &1543.75&1506.88&2.42&2951.92&3&1791.62&1542.79&1538.31&0.19&873.18&12&1543.60&1542.79&1538.12&0.20&886.52&12\\
 &    & 3 & 15 &1324.31&1323.69&0.06&920.51&14&1613.60&1324.31&1324.31&0.00&39.91&15&1328.88&1324.31&1324.31&0.00&56.76&15\\
 & 40 & 2 & 15 &2593.24&2346.40&9.69&3600.43&0&2927.91&2584.92&2458.66&4.83&3600.63&0&2576.07&2566.41&2464.83&3.88&3600.42&0\\
 &    & 3 & 15 &1817.38&1626.78&10.14&3600.49&0&2385.57&1786.19&1737.70&2.87&3600.55&0&1791.05&1783.76&1740.60&2.49&3600.39&0\\
 & 60 & 2 & 15 &11477.57&3140.18&59.11&3600.43&0&3849.91&3641.43&3323.75&8.47&3593.49&1&3622.28&3560.31&3326.08&6.39&3600.51&0\\
 &    & 5 & 15 &1944.24&1779.20&9.50&3172.28&3&3050.93&1942.50&1838.62&6.08&2891.80&3&1976.12&1928.44&1843.69&5.04&2912.93&3\\
 & 80 & 2 & 15 &12802.34&6539.05&37.60&3601.13&0&7887.26&7643.68&6807.73&11.08&3600.95&0&7811.17&7652.87&6805.83&11.17&3600.62&0\\
 &    & 5 & 15 &6506.24&2003.49&43.60&3600.54&0&3664.65&3183.77&2129.40&32.82&3600.85&0&3094.88&2862.02&2132.01&25.97&3600.78&0\\
\midrule
4& 10 & 2 & 15 &900.07&900.07&0.00&1.40&15&998.15&900.07&900.07&0.00&0.87&15&900.07&900.07&900.07&0.00&0.98&15\\
 & 20 & 2 & 15 &1474.73&1453.28&1.48&2880.62&3&1643.90&1474.32&1473.88&0.03&1102.13&13&1477.04&1474.32&1474.32&0.00&1081.49&15\\
 &    & 3 & 15 &1225.08&1215.71&0.61&1939.18&8&1550.94&1225.08&1225.08&0.00&571.73&15&1225.91&1225.08&1225.08&0.00&667.41&15\\
 & 40 & 2 & 15 &2926.95&2794.80&5.87&2802.07&6&3371.45&2908.20&2874.51&1.56&2227.70&6&2908.12&2902.08&2875.95&1.22&2274.84&6\\
 &    & 3 & 15 &1896.24&1789.37&5.96&3600.86&0&2583.84&1886.98&1813.90&3.96&3600.46&0&1887.00&1877.35&1815.24&3.46&3600.34&0\\
 & 60 & 2 & 15 &5008.15&4693.89&6.40&3602.67&0&5411.16&4941.90&4746.74&4.02&3601.14&0&4997.54&4927.10&4750.20&3.48&3600.90&0\\
 &    & 5 & 15 &2158.35&1769.79&17.12&3600.69&0&3074.79&2116.96&1849.74&13.39&3600.95&0&2097.69&2024.03&1850.70&8.95&3600.83&0\\
 & 80 & 2 & 15 &7574.95&5269.63&27.08&3600.44&0&6296.40&6016.62&5377.15&10.73&3600.93&0&6153.56&5984.27&5379.73&10.87&3600.91&0\\
 &    & 5 & 15 &2506.50&1835.31&26.21&3600.38&0&2905.83&2524.22&1922.26&22.62&3600.96&0&3141.59&2729.09&1922.24&26.89&3601.21&0\\
\midrule
Tot. &&& 540 &  115127.07 &  60847.15 &  17.37 & 2557.26 &   173 &  79520.32 & 68311.33 &  62701.55 & 5.68 & 2187.38 &  223 &  68632.15 & 67226.56 &  62771.34 &  4.10 & 2184.83 &  224 \\
\bottomrule
\end{tabularx}
    \begin{tablenotes}
    \item $U^{\mathrm{sum}}$: Sum of upper bounds, $L^{\mathrm{sum}}$: Sum of lower bounds, $G^{\mathrm{avg}}$: Avg. \acrshort{mip} Gap (in \%), $T^{\mathrm{avg}}$: Avg. runtime, $\#$o: Number of optimal solved runs, $U_0^{\mathrm{sum}}$: Sum of objective values of initial solutions.
    \end{tablenotes}
    \caption{Detailed results of new benchmark tests categorized by classes, part numbers, and machine numbers}
    \label{tab:newTestsdetailed}
        
    \end{threeparttable}
    \end{scriptsize}
\end{table}

\begin{table}[!htb]
\newcolumntype{U}{>{\raggedleft\arraybackslash}X}
\newcolumntype{T}{>{\raggedleft\arraybackslash}X}
\newcolumntype{G}{>{\raggedleft\arraybackslash}X}
\newcolumntype{O}{>{\raggedleft\arraybackslash}X}
    \centering
    \begin{scriptsize}
    \begin{threeparttable}
    \begin{tabularx}{1.36\textwidth}{lllPrrrrrrrPrrrrrrrP}
\toprule
&&&& \multicolumn{8}{c}{{\BCo}} & \multicolumn{8}{c}{{\BCts}}\\
\cmidrule(lr{0.5em}){5-12} 
\cmidrule(lr{0.5em}){13-20} 
Cl. & $\vert \mathcal{I} \vert$ & $\vert \mathcal{M} \vert$ &  {\#}runs &  $\#$E & $Cuts_{\mathrm{LB}}$ & $Cuts_{\mathrm{OR}}$ & $Cuts_{\mathrm{BR}}$ & $Cuts_{\mathrm{OP}}$ & $T_c^{\mathrm{avg}}$ & $t^{\mathrm{max}}_{\mathrm{OP}}$   & {\#}HP & $\#$E & $Cuts_{\mathrm{LB}}$ & $Cuts_{\mathrm{OR}}$ & $Cuts_{\mathrm{BR}}$ & $Cuts_{\mathrm{OP}}$ & $T_c^{\mathrm{avg}}$ & $t^{\mathrm{max}}_{\mathrm{OP}}$   & {\#}HP \\
\midrule
1&10&2& 15 &11.80&0.00&0.00&0.00&0.00&0.30&0.03&0.00&4.93&0.00&0.00&0.00&0.00&0.08&0.03&0.00\\
 &20&2& 15 &17.73&0.00&0.00&0.00&0.00&1.33&0.05&0.00&4.00&0.00&0.00&0.00&0.00&0.15&0.04&0.00\\
 &  &3& 15 &29.67&0.00&0.80&0.00&2.40&2.26&0.09&0.00&8.93&0.00&0.60&0.00&1.40&0.71&0.04&0.00\\
 &40&2& 15 &52.53&0.00&3.73&2.40&19.73&1194.65&768.02&0.80&26.33&0.00&1.67&1.53&9.47&1092.11&949.95&0.53\\
 &  &3& 15 &111.80&3.73&17.40&5.07&18.53&1181.44&978.01&0.60&79.60&3.60&21.20&13.27&20.33&677.63&538.09&0.47\\
 &60&2& 15 &454.13&0.20&178.60&20.53&152.47&1757.12&1277.19&0.87&256.60&0.13&109.00&16.53&90.20&1477.26&1083.87&0.60\\
 &  &5& 15 &155.60&0.00&40.53&2.13&60.93&211.98&104.58&0.07&69.20&0.00&21.93&0.80&35.67&201.71&122.39&0.07\\
 &80&2& 15 &225.60&0.00&93.20&4.53&54.60&2923.19&2263.82&1.53&95.27&0.00&33.00&4.27&27.47&2471.12&2142.25&1.20\\
 &  &5& 15 &222.33&0.00&95.60&10.93&85.13&2725.66&2371.09&1.27&210.20&0.00&84.60&10.73&71.33&1342.45&1015.90&0.73\\
\midrule
2&10&2& 15 &10.87&0.00&0.00&0.00&0.00&0.32&0.03&0.00&4.40&0.00&0.00&0.00&0.00&0.09&0.03&0.00\\
 &20&2& 15 &43.40&0.00&0.53&0.53&17.73&27.51&10.97&0.00&22.07&0.00&0.47&0.13&11.53&25.92&16.31&0.00\\
 &  &3& 15 &66.73&0.00&11.33&1.33&24.53&2.75&0.11&0.00&27.93&0.00&5.00&0.00&20.07&1.52&0.09&0.00\\
 &40&2& 15 &641.53&0.00&747.20&4.47&176.07&1781.60&1539.80&0.60&401.33&0.00&486.87&1.47&103.47&2033.40&1528.13&1.07\\
 &  &3& 15 &696.67&0.00&476.07&24.27&376.67&1922.57&1436.20&0.87&421.93&0.00&291.93&14.87&216.33&1499.54&1219.49&0.60\\
 &60&2& 15 &293.00&1.00&99.60&12.00&124.93&3324.43&1956.03&2.27&106.73&0.67&27.13&5.33&43.73&2973.48&2248.27&1.53\\
 &  &5& 15 &420.73&0.00&318.27&25.20&263.47&1131.48&807.89&0.67&311.80&0.00&196.87&13.40&195.60&339.39&158.22&0.33\\
 &80&2& 15 &766.87&0.00&878.93&19.53&270.00&2544.51&1757.97&1.60&515.00&0.00&636.93&17.93&172.27&2017.49&1359.91&1.40\\
 &  &5& 15 &324.27&3.20&309.67&24.33&260.80&1675.42&1423.86&0.93&262.27&4.93&224.07&12.80&219.07&1099.27&877.92&0.60\\
\midrule
3&10&2& 15 &37.87&0.00&20.67&0.00&14.20&0.48&0.03&0.00&19.13&0.00&12.53&0.00&8.73&0.23&0.03&0.00\\
 &20&2& 15 &531.73&7.20&359.53&27.47&132.93&8.30&0.06&0.00&391.73&7.20&343.60&24.53&114.93&6.28&0.07&0.00\\
 &  &3& 15 &182.67&2.93&93.67&1.93&32.20&2.09&0.05&0.00&57.60&2.00&46.00&1.60&15.20&0.70&0.03&0.00\\
 &40&2& 15 &2557.80&27.20&2627.73&82.27&494.67&91.18&0.93&0.00&1683.13&21.13&1990.07&52.00&285.53&48.71&0.84&0.00\\
 &  &3& 15 &1469.40&33.47&1288.20&45.53&363.60&47.76&1.24&0.00&743.13&20.40&749.40&14.20&149.40&18.96&0.53&0.00\\
 &60&2& 15 &1021.40&14.13&765.93&41.60&189.93&224.08&174.99&0.07&957.33&11.20&750.80&42.93&156.27&48.55&13.81&0.00\\
 &  &5& 15 &911.40&0.00&1425.13&35.73&147.27&10.57&0.09&0.00&634.20&0.00&1044.53&11.20&71.80&6.57&0.08&0.00\\
 &80&2& 15 &820.80&0.00&1144.93&15.87&140.87&72.33&14.35&0.00&695.27&0.00&946.40&10.00&102.80&234.03&162.26&0.13\\
 &  &5& 15 &569.13&1.20&1508.73&14.20&51.40&8.33&0.15&0.00&488.40&0.40&1257.93&10.07&42.27&7.17&0.14&0.00\\
\midrule
4&10&2& 15 &24.07&0.07&3.93&0.00&1.40&0.37&0.03&0.00&10.20&0.00&2.20&0.00&0.80&0.12&0.02&0.00\\
 &20&2& 15 &352.87&0.00&319.40&5.07&77.73&7.45&0.21&0.00&289.33&0.00&306.20&2.53&52.27&4.49&0.05&0.00\\
 &  &3& 15 &212.60&1.73&96.80&11.87&11.80&2.02&0.03&0.00&137.53&1.33&73.47&12.40&7.60&1.06&0.03&0.00\\
 &40&2& 15 &1598.67&53.53&1056.27&13.80&113.20&36.56&0.33&0.00&983.40&36.67&708.47&6.13&44.33&15.61&0.26&0.00\\
 &  &3& 15 &1467.60&5.87&2135.27&22.87&312.40&51.87&0.82&0.00&690.33&2.60&1225.60&8.87&122.93&18.36&0.39&0.00\\
 &60&2& 15 &2221.27&0.00&2985.60&17.47&285.73&83.09&0.77&0.00&1478.27&0.00&2129.07&10.73&148.33&42.25&0.57&0.00\\
 &  &5& 15 &1171.20&12.53&2062.67&15.13&143.73&21.91&0.37&0.00&863.80&8.53&1322.00&9.60&71.80&11.32&0.27&0.00\\
 &80&2& 15 &1075.87&23.33&1450.60&19.40&151.20&176.76&91.03&0.13&842.33&17.27&895.53&12.13&112.87&248.23&208.65&0.13\\
 &  &5& 15 &515.87&9.40&699.60&2.87&54.00&9.15&1.13&0.00&439.07&5.60&596.80&4.40&42.47&6.77&0.19&0.00\\
\bottomrule
\end{tabularx}
    \begin{tablenotes}
        \item $\#$E: Number of batches checked, $Cuts_{\mathrm{LB}}$, $Cuts_{\mathrm{OR}}$, $Cuts_{\mathrm{BR}}$, $Cuts_{\mathrm{OP}}$: Avg. number of cuts, $T^{avg}$: Avg. runtime, $T_c^{\mathrm{avg}}$: Avg. check time, {\#}T/O: Number of runs with timeout within orthogonal packing.
    \end{tablenotes}
    \caption{Cut analysis of \acrshort{bnc} approaches categorized by classes and part numbers}
    \label{tab:newTestsCuts}
    \end{threeparttable}
    \end{scriptsize}
\end{table}
\end{landscape}
}

\section{Conclusion}
\label{sec:C}
This study is motivated by the planning processes in additive manufacturing, where several parts can be simultaneously processed together in the same batch on a 3D printer. We investigate the integrated problem of two-dimensional orthogonal packing with rotation and scheduling on unrelated parallel machines. In order to solve this problem exactly, we propose a new branch-and-cut approach, in which we use and extend several adaptions of state-of-the-art procedures and techniques from the packing literature. In extensive computational tests on several data sets, we assess two variants of the proposed branch-and-cut and demonstrate their superior algorithmic performance compared to an integrated mixed-integer programming model taken from the literature. In addition, we analyze the structure of existing test data for scheduling in additive manufacturing and propose a new benchmark test set. Using this new benchmark set, we show that the size of parts greatly impacts whether the scheduling problem or the packing problem is more prominent. In a comprehensive evaluation of the test results, we find that using dual feasible functions can effectively prove infeasibility of packing sub-problems. This is especially the case in comparison with lower bound techniques, which only add significant value in the solution process for instances with larger parts. In summary, the considered stepwise procedure prevented the algorithm from invoking other feasibility checks, which are more costly in terms of computational effort. 

The proposed algorithms struggle to determine feasibility or infeasibility for packing sub-problems with many parts that are small relative to the build area. To improve early infeasibility detection, reduction techniques that exist for the packing problem variant with fixed orientation should be extended to the case with rotation. For practitioners, several problem extensions must be explored in order to improve real-world applicability. For example, considering minimum distances between processed parts enables ensuring a certain level of printing quality. Considering irregularly-shaped parts enables increases in build space utilization. Similarly, extending the sub-problem to a 3D packing problem enables higher build space utilization for additive manufacturing technologies that allow the stacking of parts. Furthermore, the considered solution procedures may be applied to other related problems -- for example, the two-dimensional bin packing problem with rotation or the two-dimensional orthogonal packing problem with rotation. Further studies should show the extent to which incorporating these methods can improve the existing benchmark results.



\afterpage{
\appendix
\section{Details on Machine Parameters}
\label{MGD}
The following \Autoref{tab:machineGenerationDetails} presents the considered specifications and formulas, which are used to define the machine types and the individual parameters of each machine.

\renewcommand{\arraystretch}{1.05}
\begin{table}[htb]
\begin{footnotesize}
\begin{tabularx}{\textwidth}{lRRR}
    \toprule
    \multicolumn{4}{l}{Machine Build Spaces}\\
    \midrule
    & $\Omega$ & $\Lambda$ & $H$ \\
    \midrule
		Machine type 1 & 40 & 40 & 25 \\
		Machine type 2 & 25 & 25 & 20 \\
		Machine type 3 & 28 & 28 & 24 \\
		Machine type 4 & 28 & 28 & 32 \\
		Machine type 5 & 40 & 40 & 20 \\
		Machine type 6 & 28 & 50 & 36 \\
	\toprule
	\multicolumn{3}{l}{Determination of processing times}\\
	\midrule
	Scan speed $\sigma$ & \multicolumn{3}{l}{uniformly random integers in [8, 11] $\cdot 1,000 \frac{mm}{s}$}\\
	Layer production speed $\phi$ & \multicolumn{3}{l}{uniformly random integers in [3, 7] $\frac{s}{\mathrm{layer}}$}\\
	Laser diameter $\delta$ & \multicolumn{3}{l}{uniformly random integers in [8, 11] $\cdot \frac{1}{100} mm$}\\
	Layer thickness $\theta$ & \multicolumn{3}{l}{uniformly random integers in [4, 10] $\cdot \frac{1}{100} mm$}\\
    Setup time $T^S_m$ & \multicolumn{3}{l}{uniformly random in [1, 2] $h$}\\
    Scanning time $T^L_m$ & \multicolumn{3}{l}{$T^L_m = \frac{1}{\sigma \cdot \delta \cdot \theta}$ $\frac{h}{\mathrm{cm^3}}$}\\
    Recoating time $T^R_m$ & \multicolumn{3}{l}{$T^R_m = \frac{\phi}{\theta}$ $\frac{h}{\mathrm{cm}}$}\\
    \bottomrule
	\end{tabularx}
	\end{footnotesize}
	\caption{Definition of machine parameters}
	\label{tab:machineGenerationDetails}
\end{table}

\printcredits

\bibliographystyle{cas-model2-names}

\bibliography{IPAM_bib}
}
\bio{}
\endbio

\bio{}
\endbio

\end{document}